\documentclass[12pt,a4paper]{article}

\usepackage{latexsym}
\usepackage{amsmath}
\usepackage{amsfonts}
\usepackage{amssymb}

\newcommand{\be}{\begin{equation}}                                          %
\newcommand{\ee}{\end{equation}}                                            %
\newcommand{\bea}{\begin{eqnarray}}                                         %
\newcommand{\eea}{\end{eqnarray}}                                           %
 \renewcommand{\theequation}{\arabic{section}.\arabic{equation}}            %
\def\tdl{{{\delta}_l}}                                                      %
\def\tdr{{{\delta}_r}}                                                      %
 \def\td{{{\delta}}}                                                        %
 \def\G{{\cal G}}                                                           %
 \def\cN{{\cal N}}                                                          %
  \def\I{{\cal I}}                                                          %
  \def\H{{\cal H}}                                                          %
  \def\B{{\cal B}}                                                          %
  \def\A{{\cal A}}                                                          %
  \def\Z{{\cal Z}}                                                          %
 \def\cC{{\cal C}}                                                          %
 \def\ri{{\mathrm{i}}}                                                      %
  \def\cR{{\cal R}}                                                         %
  \def\cL{{\cal L}}                                                         %
  \def\bR{{\mathbb R}}                                                      %
  \def\bC{{\mathbb C}}                                                      %
  \def\bT{{\mathbb T}}                                                      %
 \def\1{{\mbox{\boldmath $1$}}}                                             %
  \def\tr{\mathrm{tr\,}}                                                    %
  \def\diag{\mathrm{diag}}                                                  %
    \def\w{\wedge}                                                          %
  \def\hp{\hat P_c}                                                         %
  \def\ho{\hat\omega_c}                                                     %
  \def\fp{\frac{x}{2}}                                                      %
  \def\fv{\frac{\vert x\vert}{2}}                                           %
  \def\up{\Upsilon}                                                         %
  \def\D{{\cal D}}                                                          %
    \def\ds{\left.\frac{d}{ds}\right\vert_{s=0}}                            %
   \def\half{{\frac{1}{2}}}                                                 %
   \def\bD{{\mathbf D}}                                                     %
   \def\End{{\mathrm{End}}}                                                 %
  \def\Dress{{\mathrm{Dress}}}                                              %
 \def\jp{{\frac{1}{2}}}                                                     %
\def\js{{\frac{1}{4}}}                                                      %
 \def\trl{\triangleright}                                                   %
 \def\om{\omega}                                                            %
   \def\lm{\lambda}                                                         %
 \def\gln{GL(n,\bC)}                                                        %
\def\C{{\mathfrak C}}                                                       %
\def\bZ{{\mathbb Z}}                                                        %
\def\barG{{\bar G}}                                                         %
\def\barB{{\bar B}}                                                         %
 \def\cU{{\cal U}}                                                          %
\def\cJ{{\cal J}}                                                           %
\def\cA{{\cal A}}                                                           
\begin{document}

\vspace*{0.5cm}
\begin{center}
{\Large \bf Poisson-Lie interpretation  of trigonometric  Ruijsenaars
duality}

\end{center}

\vspace{0.2cm}

\begin{center}

L. Feh\'er${}^{a}$ and  C. Klim\v c\'\i k${}^b$ \\

\bigskip

${}^a$Department of Theoretical Physics, MTA  KFKI RMKI\\
1525 Budapest 114, P.O.B. 49,  Hungary, and\\
Department of Theoretical Physics, University of Szeged\\
Tisza Lajos krt 84-86, H-6720 Szeged, Hungary\\
e-mail: lfeher@rmki.kfki.hu

${}^b$Institut de math\'ematiques de Luminy,
 \\ 163, Avenue de Luminy, \\ 13288 Marseille, France\\
 e-mail: klimcik@iml.univ-mrs.fr

\bigskip

\end{center}

\vspace{0.2cm}

\begin{abstract}
A geometric interpretation of the duality between two real forms of the complex
trigonometric Ruijsenaars-Schneider system is presented.
The phase spaces of the systems in duality are viewed as two different models
of the same reduced phase space arising from a suitable symplectic reduction
of the standard Heisenberg double of $U(n)$.
 The collections of  commuting Hamiltonians of the systems in duality
  are shown to descend from two   families of `free' Hamiltonians
  on the double which are dual to each other in a Poisson-Lie sense.
  Our results  give  rise to
   a major simplification of Ruijsenaars'
proof of the crucial symplectomorphism property of the duality map.
\end{abstract}

\newpage

\sloppy \raggedbottom
\section{ Introduction}
\setcounter{equation}{0}

In 1986 Ruijsenaars and Schneider \cite{RS} introduced  a remarkable
deformation of the non-relativistic integrable many-body systems due
to  Calogero \cite{Cal}, Sutherland \cite{Sut}
 and Moser \cite{Mos}.
The deformation corresponds to a passage from Galilei to  Poincare
invariance, and for this reason the deformed  systems can be called
relativistic Calogero  systems. The family of Calogero type systems
is very important both from the physical and from the mathematical
point of view, and has been the subject of intense studies ever
since its inception. See, e.g., the reviews
\cite{SR-CRM,beauty,Eti}.

When constructing action-angle maps for the classical, non-elliptic
$A_{n-1}$ systems, Ruijsenaars \cite{SR-CMP,RIMS94,RIMS95}
discovered an intriguing relation that arranges the Calogero type
systems into `dual pairs'. The main feature of the duality between
system (i) and system (ii) is the fact that the action variables of
system (i) are the particle-position variables of system (ii), and
vice versa. The simplest example is provided by the non-relativistic
rational Calogero system, whose self-duality  was already noticed by
Kazhdan, Kostant, and Sternberg \cite{KKS} when treating the system
by symplectic reduction  of $T^* u(n)\simeq  u(n) \times u(n)$, and
thereby relating the symmetry between the $u(n)$ factors to the
self-duality property. In his papers Ruijsenaars hinted at the
possibility that there might exist an analogous geometric picture
behind the duality in the other cases, too, leaving this as a
problem for future investigation.

Later  Gorsky and Nekrasov \cite{GN} and their  coworkers \cite{
Fock+, FR, Nekr} introduced new ideas and conjectures in the area of
Ruijsenaars' duality. By generalizing \cite{KKS}, they proposed to
interpret the duality in general in terms of symplectic reduction of
two distinguished families of commuting Hamiltonians living on a
suitable higher dimensional phase space.
 Upon a single reduction of this phase space described using two
alternative  gauge slices, i.e., two alternative models of the
reduced phase space,
 those two families of commuting Hamiltonians
 may reduce to  the two respective sets of  action variables  of  the mutually  dual  systems.
 They  mainly focused on  reductions of  infinite-dimensional  phase spaces
 aiming to relate the Calogero type many-body  systems to field theories.
 We wish to stress that  also relatively simple  {\it finite-dimensional}
 phase spaces can be considered for which these ideas  work
 fully as expected.  For example,
 we worked out in \cite{FKinJPA}  the duality between the hyperbolic Sutherland
and the rational Ruijsenaars-Schneider systems \cite{SR-CMP} by
reducing the cotangent bundle of the group $GL(n,\bC)$.

  The key   problem in the  reduction approach   is to find for  each particular case of  Ruijsenaars'
  duality two distinguished
families of commuting Hamiltonians  on an appropriate
higher-dimensional phase space and   to find  a way   how to reduce
those families.
 In the present paper, we  solve this problem for one of the structurally
 most interesting and technically most
 involved   cases  of the duality  that  links  together two particular real forms
 of the complex trigonometric
 Ruijsenaars-Schneider  system \cite{RIMS95}.  The first  real form is
the original trigonometric Ruijsenaars-Schneider system \cite{RS}
characterized  by the Lax matrix $L$  and symplectic form $\omega$
\be L_{j,k}(q,p) =  \frac{ e^{p_k}\sinh(\frac{x}{2})}{\sinh(\ri q_j
- \ri q_k +\frac{x}{2})} \prod_{m\neq j} \left[1 +
\frac{\sinh^2\frac{x}{2}} { \sin^2(q_j - q_m)}\right]^{\frac{1}{4}}
\prod_{m\neq k} \left[1 + \frac{\sinh^2\frac{x}{2}} { \sin^2(q_k -
q_m)}\right]^{\frac{1}{4}}, \label{1.1}\ee \be \omega=\sum_{k=1}^n
dp_k\wedge dq_k,  \qquad  0\leq q_k < \pi, \quad q_1> q_2> ...> q_n,
\qquad  p_k\in \bR, \label{1.2}\ee and the other real form is  the
Ruijsenaars dual of (1.1-2)    that
 can be \emph{locally}
 characterized by the  Lax matrix $\hat L$ and
symplectic form $\hat\omega$ \be \hat L_{j,k}(\hat q,\hat p) =
\frac{e^{\ri \hat q_k}\sinh(-\frac{x}{2})}{\sinh(\hat p_j - \hat
p_k-\frac{x}{2})} \prod_{m\neq j} \left[1-
\frac{\sinh^2\frac{x}{2}}{\sinh^2(\hat p_j - \hat
p_m)}\right]^{\frac{1}{4}}\prod_{m\neq k} \left[1-
\frac{\sinh^2\frac{x}{2}}{\sinh^2(\hat p_k - \hat
p_m)}\right]^{\frac{1}{4}}, \label{1.3}\ee \be
\hat\omega=\sum_{k=1}^n d\hat p_k\wedge d\hat q_k, \qquad
 0\leq \hat q_k < 2\pi,
 \qquad \hat p_j-\hat p_{j+1}>\frac{\vert x\vert}{2}
 \quad (j=1,..., n-1). \label{1.4}\ee
The Lax matrices are generating functions for Hamiltonians in
involution. The `main Hamiltonians' from which these systems derive
their names read \be H_{\mathrm{RS}}(q,p) = \frac{1}{2}\tr\bigl(L(q,p)
+ L(q,p)^{-1}\bigr)= \sum_{j=1}^n (\cosh p_j) \prod_{m\neq
j}\biggl[1 + \frac{\sinh^2\frac{x}{2}} { \sin^2(q_j - q_m)}
\biggr]^{\frac{1}{2}} \label{1.5} \ee and \be \hat
H_{\mathrm{RS}}(\hat q,\hat p) = \frac{1}{2} \tr\bigl(\hat L(\hat
q,\hat p) + \hat L(\hat q,\hat p)^{-1}\bigr)= \sum_{j=1}^n (\cos
\hat q_j) \prod_{m\neq j}\biggl[1 - \frac{\sinh^2\frac{x}{2}} {
\sinh^2(\hat p_j - \hat p_m)}  \biggr]^{\frac{1}{2}}. \label{1.6}
\ee
 Here $x$ is a real, non-zero
coupling constant and the `velocity of light' has been set to unity.
The  variables $(q,p)$ and $(\hat q,\hat p)$
  provide  one-to-one set theoretic
parametrizations of the phase spaces $P$ and $\hat P$ of the
original and the dual Ruijsenaars-Schneider systems, respectively.
As manifolds, $P$ is the cotangent bundle of the configuration space
consisting of unordered $n$-tuples of distinct points on the circle
$U(1)$, which corresponds to $n$ indistinguishable particles moving
on the circle,  and $\hat P$ is an open submanifold of the cotangent
bundle of the torus $\bT_n = U(1)^{\times n}$. In what follows
$(P,\omega,L)$ denotes the collection
  of the Ruijsenaars-Schneider Hamiltonian
systems   defined by the spectral-invariants of $L$ (\ref{1.1}), and
similarly for $(\hat P,\hat \omega,\hat L)$.

 The duality between   the  real forms  (1.1-2) and  (1.3-4) was previously studied in
\cite{RIMS95} by the `direct' approach, or, in other words, by  an
approach that did not use the methods of symplectic reduction. In
particular, Ruijsenaars presented integration algorithms for the
Hamiltonian flows of the original and dual  systems
 and proved that the   flows of the original   system are  complete on $(P,\omega)$  but the flows
 of the dual system are not complete on  $(\hat P, \hat \omega)$.  He also pointed out that
 this singular behaviour of the dual system is related to the fact  that  the injective action-angle map  from
$(\hat P, \hat \omega)$  into $(P,\omega)$ is not surjective but
has only  a dense open image. He then introduced  an extension
$(\hat P_c, \hat \omega_c)$  of the   phase space   $(\hat P, \hat
\omega)$
 in order to achieve   bijectivity of the corresponding extension of the
 action-angle map (alias the duality map)\footnote{Our notations $P$, $\hat P$, $\hat P_c$ correspond,
 respectively, to $\Omega$, $\hat \Omega$, $\hat \Omega^\sharp$ in
\cite{RIMS95}. Further notational correspondence is given in
footnotes 4
 and 5 in the  text.}.
 Remarkably, the dual flows turned out to be complete
 on   the extended phase space  $(\hat P_c, \hat \omega_c)$, which can be therefore
 referred to as the {\it completion} of  $(\hat P, \hat \omega)$.

 Our interest in the real forms of the complex trigonometric Ruijsenaars-Schneider system
 was inspired by a conjecture that
Gorsky and Nekrasov raised in \cite{GN}. They derived yet another
trigonometric real form
  (called the $\mathrm{III}_{\mathrm{b}}$ system  in \cite{RIMS95})  from gauged WZW theory,
  and conjectured that it should be   possible to derive the same system  also
 from an appropriate finite-dimensional Heisenberg double. Although, as it stands, this
conjecture remains still open, we recently succeeded to prove its
`extrapolation' to  the trigonometric real form (1.1-2).    Indeed,
in
 our paper  \cite{LMP},  we  obtained
the collection $(P, \omega,L)$ of the  original
Ruijsenaars-Schneider Hamiltonian systems  by a reduction of a
certain family  of `free' Hamiltonians on the Heisenberg double of
the standard compact Poisson-Lie group  $U(n)$. The `free'
Hamiltonians were constructed as the pull-backs of the
dressing-invariant functions on the dual Poisson-Lie group
 $B(n)$ by the Iwasawa map $\Lambda_R$.
(See Section 2.1 for the definitions of the Iwasawa maps and of the
pertinent  actions of $U(n)$.) Their basic properties
 are explicit integrability and invariance
with respect to the so-called quasi-adjoint action of $U(n)$ on the
 double.
Our present investigation relies on the existence of a dual family
of `free'
 Hamiltonians that have the same
properties. In fact, the members of the dual family can be
constructed by
 pulling-back to the double the conjugation-invariant
 functions on the Poisson-Lie group $U(n)$ by means of the dual Iwasawa
 map $\Xi_R$.
     As the main technical result of the present  paper, {\it we shall  obtain the
completion  of the  collection   $(\hat P,\hat\omega,\hat L)$
 of the dual
Ruijsenaars-Schneider Hamiltonian systems by reducing the dual
family of free Hamiltonians on the Heisenberg double of   $U(n)$}.
Notice  that the first occurrence of the term  `dual'  in the
previous sentence refers to  Ruijsenaars' duality  of integrable
systems and the second  occurrence refers to duality  in the sense
of   Poisson-Lie groups. Said in other words,
 the  duality  between the two trigonometric Ruijsenaars-Schneider systems
{described in \cite{RIMS95}} can be interpreted as a remnant of the
`geometric democracy' between the Poisson-Lie group $U(n)$ and  the
corresponding   dual Poisson-Lie group $B(n)$ that survives the
symplectic reduction. We believe that our results  are related to
the quantum group aspects of the standard trigonometric {
Ruijsenaars-Schneider}  system \cite{SR-Kup,Eti,Noum,Mim} through
the general correspondence between quantum groups and Poisson-Lie
groups.

Both the original and the dual   families
  of free Hamiltonian systems  on the Heisenberg double are Poisson-Lie symmetric with respect
to the   quasi-adjoint action  of $U(n)$ on the double. A  decisive
step in  our work  \cite{LMP} was the choice of  a suitable value
$\nu(x)$ of the Poisson-Lie moment map $\Lambda$ of this
quasi-adjoint symmetry such
  that  the original Ruijsenaars-Schneider phase space $(P,\omega)$  could be identified with
 the   constraint-manifold $F_{\nu(x)}:=\Lambda^{-1}(\nu(x))$ factorized by the
 gauge group given by the isotropy  group $G_{\nu(x)}< U(n)$.
In other words, $(P, \omega)$ was identified in \cite{LMP} with the
reduced phase space arising from the symplectic reduction of the
Heisenberg double by the quasi-adjoint Poisson-Lie symmetry at the
value $\nu(x)$ of the moment map $\Lambda$. In this paper, we shall
demonstrate that  the completion $(\hat P_c,\hat\omega_c)$ of the
dual phase space $(\hat P,\hat\omega)$ is also symplectomorphic to
the  {\it same} reduced phase space $F_\nu(x)/G_\nu(x)$. Therefore
we can view $(P,\omega)$ and $(\hat P_c, \hat \omega_c)$ as two
distinct models of a single reduced phase space.
 Moreover, as was already mentioned, we shall interpret the two collections
   of commuting Hamiltonians associated with the dual pair of Ruijsenaars-Schneider
  systems as reductions of two commutative families of free Hamiltonians
  on the Heisenberg double.
Our results thus  fit the geometric interpretation of Ruijsenaars'
duality advocated by Gorsky and his
collaborators for example in \cite{Fock+}.

Speaking generally, { the} symplectic reduction approach often
represents { useful} technical streamlining or simplification with
respect to direct methods. { This feature occurs also in our
particular case.}  For { instance}, the free Hamiltonian    systems
on the Heisenberg  double { are} themselves integrable and, as we
shall see,  their very simple Lax matrices   reduce to {the}
relatively complicated Lax matrices of the Ruijsenaars-Schneider
systems. In addition, we { can recover the integration algorithms}
for the original and  dual Ruijsenaars-Schneider flows \cite{RIMS95}
{simply by}  projecting   the   `free' flows on the respective
models of the reduced phase space  $F_{\nu(x)}/G_{\nu(x)}$,  {and
these projected flows are automatically complete}.
 However,
 there is one aspect of the duality story where the reduction approach yields more than
technical streamlining or simplification  and, in fact, represents
a {\it  major technical advantage} with respect to the direct
approach. Indeed, in the direct approach  it was very difficult  to
prove  the crucial symplectomorphism property of the extended
action-angle  map between the phase space $(P,\omega)$ and the
extended dual phase space $(\hat P_c, \hat \omega_c)$.  To realize
the difficulties, the reader may consult   the  `hyperbolic   proof'
published in   \cite{SR-CMP} and   its  `trigonometric analytic
continuation' presented in \cite{RIMS95}.   In these references, a
sophisticated web of non-trivial steps  had been used  combining
  scattering theory with
  demanding analysis  and intricate analytic continuation
  arguments.
  However, from the point of view of  the reduction approach the extended action-angle map  is
 automatically a symplectomorphism, since it is easily  recognized to be the composition of
 the symplectomorhism relating
 $(P,\om)$ with $F_{\nu(x)}/G_{\nu(x)}$ and of the symplectomorphism relating
 $F_{\nu(x)}/G_{\nu(x)}$ with $(\hat P_c,\hat \omega_c)$.

 To render justice to the direct methods, we note that so far considerably
  more examples of the duality were   thoroughly investigated
in the direct approach \cite{SR-CMP,RIMS94,RIMS95}  than in the
symplectic reduction approach. We hope, however,  that the
attractive features mentioned above supply sufficient motivation to
further develop the reduction approach, along the lines discussed at
the end of this article.

The  rest of the paper is organized as follows. In Section 2, we
first review the geometry of the standard Heisenberg double of
$U(n)$ and recall the definition of the quasi-adjoint action
of $U(n)$ on it. We then describe the two    families of
 free Hamiltonian systems  on the double
that are Poisson-Lie symmetric with respect to the quasi-adjoint
action. We do not claim any new results in this section, although we
could not find in  the literature any  previous detailed treatment
of the free flows as given by our Propositions 2.1 and 2.2.

The reduction of the two   families of free
 Hamiltonians to the respective two real forms of the complex  trigonometric
Ruijsenaars-Schneider system is presented in Section 3. After a
summary of required notions, Theorems 3.1 and 3.2 state in a
strengthened form the result of \cite{LMP},  where the first family
was reduced to  the system $(P,\omega,L)$. Then the new Theorems 3.3
and 3.4 are formulated, which claim
 that the same symplectic reduction applied
to the second family  yields  the completion
 of the system $(\hat P, \hat \omega, \hat L)$.
 Section 4 is devoted to the proofs of Theorems 3.1 and 3.2.  Although many results described in this section
 were obtained already in our letter \cite{LMP},  we still add here important new material. Namely, we
 pay special attention to  the Weyl group covering of the phase space $P$
  of the original Ruijsenaars-Schneider
 system,  since this enables us to recover the geometric meaning of
 the standard  coordinates on $P$ from the reduction.
 Indeed, in \cite{LMP}, we identified the Ruijsenaars-Schneider system with the
  reduced system on $F_{\nu(x)}/G_{\nu(x)}$
 by means of a somewhat mysterious coordinate transformation that was just `cooked up'
   to do the job. Here we explain the geometrical origin of this transformation.

Theorems 3.3 and 3.4 represent the main results obtained in this
paper and their proof occupies  much of Section 5. As a by-product,
the integration algorithm for the dual flows is also treated at the
end of Section 5. Further discussion, including comparison with
\cite{RIMS95}, is offered in Section 6. Finally, Appendix A
  contains an alternative proof of the fact that
  $F_{\nu(x)}$ is an
  embedded submanifold of the
  Heisenberg double, and Appendix B is devoted to the subtle
  topology of the
  configuration space of $n$ indistinguishable point-particles  moving on the circle.

\setcounter{equation}{0}

\section{ Free systems on the Heisenberg double}

Let us recall that a Poisson-Lie group  is a Lie group, $G$,  equipped with
a Poisson  bracket, $\{.,.\}_G$,
such that the multiplication map $G\times G\to G$ is Poisson.  It is an important
concept that makes it possible
to generalize the usual notion of  symmetry for Hamiltonian systems.
Namely, if the Poisson-Lie group $G$
acts on a symplectic manifold $M$ in a Poisson way (i.e.~the action map
$G\times M\to M$
is Poisson) and, moreover, if the Hamiltonian  $H$ is $G$-invariant, then
one says that
the system $(M,H)$ is $G$ Poisson-Lie symmetric \cite{ST}. For our purpose we shall
focus on the group $G=U(n)$ equipped with
its standard Poisson-Lie structure,
although all results collected in this section hold true, with minor
modifications,
for any compact reductive group $G$.

  In  subsection 2.1, we summarize the necessary information
concerning the structure of the Heisenberg double of the
dual pair of Poisson-Lie groups $U(n)$ and $B(n)$ and also recall the notion
of the quasi-adjoint action of $U(n)$ on the double.
Then, in subsection 2.2,  we describe the two families of $U(n)$ Poisson-Lie
symmetric Hamiltonian systems that
will  descend upon symplectic reduction to the trigonometric
Ruijsenaars-Schneider system (\ref{1.1})
and its dual.
  These systems turn out to be explicitly integrable, and for this reason
   we call the underlying Hamiltonians `free' Hamiltonians.

\subsection{Recall of the Heisenberg double and the quasi-adjoint action}

The Heisenberg double  is a Poisson manifold $(D, \{.,.\}_+)$ that Semenov-Tian-Shansky \cite{ST}
associated with any Poisson-Lie group $(G, \{.,.\}_G)$.  The manifold $D$ is itself
a Lie group, the Drinfeld double of $G$, and its Poisson bracket
$\{. ,.\}_+$
can be expressed in terms of the factorizable $r$-matrix of $\mathrm{Lie}(D)$.
One can also directly define the double $(D, \{.,.\}_+)$ and then recover
from it the Poisson-Lie group $(G, \{.,.\}_G)$ and its dual.

Consider the \emph{real} Lie group $D:= GL(n,\bC)$ and endow the corresponding real Lie algebra
$\D:= gl(n,\bC)$ with the non-degenerate, invariant `scalar product'
\be
(X,Y)_\D:=\Im \tr(XY),
\qquad
\forall X,Y\in \D.
\label{2.1}\ee
Here  $\Im z$ stands for the imaginary part of the complex number $z$.
Let $B:=B(n)$ be the subgroup of $D$
formed by the
upper-triangular matrices having positive entries along the diagonal,
and denote $G:= U(n)$.
As a real vector space, we  have the direct sum decomposition
\be
\D = \G + \B,
\label{2.2}\ee
where $\G:= \mathrm{Lie}(G)=u(n)$ and  $\B:= \mathrm{Lie}(B)$   are isotropic subalgebras mutually dual
to each other with respect to the pairing provided by $(.,.)_\D$.
In other words, we have a so-called Manin triple in our hands, and thus
the real Lie group $D$ carries two natural Poisson structures
$\{.,. \}_\pm$.
Here we need only the structure $\{.,.\}_+$, and
to define it we introduce the projection operators
$\pi_\G: \D \to \G$ and $\pi_\B: \D\to \B$
associated with the splitting (\ref{2.2}).
Furthermore, for any real function $\Phi \in C^\infty(D)$ introduce
the left- and right gradients $\nabla^{L,R} \Phi\in C^\infty(D,\D)$ by
\be
\ds \Phi(e^{sX} K e^{sY}) = (X, \nabla^L \Phi(K))_\D +
(Y,\nabla^R \Phi(K))_\D,
\qquad
\forall X,Y\in \D,\, \forall K\in D.
\label{2.3}\ee
By using the factorizable $r$-matrix of $\D$,
\be
\rho:= \half (\pi_\G - \pi_\B),
\label{2.4}\ee
for any $\Phi_1, \Phi_2\in C^\infty(D)$ one has  the Poisson bracket
\be
\{ \Phi_1, \Phi_2\}_+(K)=  \left( \nabla^R \Phi_1(K), \rho(\nabla^R\Phi_2(K))\right)_\D
+\left( \nabla^L \Phi_1(K), \rho(\nabla^L\Phi_2(K))\right)_\D.
\label{2.5}\ee
The Poisson manifold $(D,\{. ,. \}_+)$ is called the Heisenberg double of $G$.

By the Iwasawa decomposition,  each  element $K\in D$ has the unique representations
\be
K= b_L g_R^{-1}
\quad\hbox{and}\quad
K= g_L b_R^{-1}
\quad\hbox{with}\quad
b_{L,R}\in B,\,\, g_{L,R}\in G.
\label{2.6}\ee
As a result of the global character of this decomposition,
our $(D,\{.,.\}_+)$ is actually symplectic.
The underlying symplectic form, called $\omega_+$, was found  by Alekseev and Malkin \cite{AM}.
In fact, with the help of   the Iwasawa maps
$\Lambda_{L,R}: D \to B$ and
$\Xi_{L,R}: D \to G$ defined by using (\ref{2.6}) as
\be
\Lambda_{L,R}(K):= b_{L,R}
\quad\hbox{and}\quad
\Xi_{L,R}(K):= g_{L,R},
\label{Iwasawa}\ee
 one has
 \be
  \om_{+}=\jp \Im \tr(d\Lambda_L\Lambda_L^{-1}\wedge d\Xi_L\Xi_L^{-1})+
  \jp\Im \tr(d\Lambda_R\Lambda_R^{-1}\wedge d\Xi_R\Xi_R^{-1}).
  \label{ST}\ee

Having described the Heisenberg double, next we recall how the Poisson bracket $\{.,.\}_+$
induces Poisson-Lie structures on the groups $B$ and $G$.
 As a preparation,
 define the left- and right derivatives
$d^{L,R}f \in C^\infty(B,\G)$ for a real function $f\in C^\infty(B)$  by the equality
\be
\ds f(e^{sX} b e^{sY}) = \left(X, d^L f(b)\right)_\D +
\left(Y, d^R f(b)\right)_\D,
\qquad
\forall X,Y\in \B, \, \forall b\in B.
\label{2.9}\ee
For a real function $\phi \in C^\infty(G)$  define
$d^{L,R}\phi \in C^\infty(G,\B)$ similarly,
\be
\ds \phi(e^{sX} g e^{sY}) = \left(X, d^L \phi(g)\right)_\D +
\left(Y, d^R \phi(g)\right)_\D,
\qquad
\forall X,Y\in \G, \, \forall g\in G.
\label{2.10}\ee
By using the negative-definite scalar product on $\G$ furnished by
\be
\langle X, Y\rangle_\G := \tr(XY),
\qquad
\forall X,Y\in \G,
\label{2.11}\ee
introduce also $\bD^{L,R}\phi \in C^\infty(G, \G)$ by
\be
\ds \phi(e^{sX} g e^{sY}) = \langle X, \bD^L \phi(g)\rangle_\G +
\langle Y, \bD^R \phi(g)\rangle_\G,
\qquad
\forall X,Y\in \G, \,\forall g\in G.
\label{2.12}\ee
Defining
$R^\ri\in \End(\G)$ as
\be
R^\ri(X) = \pi_\G(-\ri X), \qquad
\forall X\in \G,
\label{2.13}\ee
which is actually nothing but the standard $r$-matrix of $\G$,
it is easy to check that the two types of derivatives over $G$ are related by
\be
d^L \phi = \ri \bD^L \phi + R^\ri (\bD^L \phi),
\qquad
d^R \phi = \ri \bD^R \phi + R^\ri (\bD^R \phi),
\label{2.14}\ee
which in particular implies that
\be
d^{L,R} \phi = \pi_\B(\ri \bD^{L,R} \phi),
\qquad \forall \phi\in C^\infty(G).
\label{2.15}\ee

After fixing all the above notations, we are ready to write down explicit formulas
for  the Poisson-Lie structures $\{.,.\}_B$ on $B$ and $\{.,.\}_G$ on  $G$. In fact,
it can be shown  that the  algebras of functions $C^\infty(B)$ and $C^\infty(G)$  pulled-back,
respectively, by the Iwasawa maps $\Lambda_R$ and $\Xi_R$  form two Poisson subalgebras
of $C^\infty(D)$.  Because of the
  surjectivity of the Iwasawa maps in (\ref{Iwasawa}), the bracket $\{.,.\}_+$ on $D$  thus  induces
  two brackets  $\{.,.\}_B$ and $\{.,.\}_G$  by the following prescriptions:
  \be
\{\Lambda_R^* f_1, \Lambda_R^* f_2\}_{+}= \Lambda_R^* \{ f_1, f_2\}_B
\quad\hbox{and}\quad
\{\Xi_R^* \phi_1, \Xi_R^* \phi_2\}_{+}= \Xi_R^* \{ \phi_1, \phi_2\}_G.
\label{2.16}\ee
The brackets $\{.,.\}_B$ and $\{.,.\}_G$ induced in this way are just the Poisson-Lie brackets on
$B$ and on $G$, respectively.
The explicit form of the induced Poisson bracket on $B$ reads
\be
\{ f_1, f_2\}_B(b)= -  \left( b^{-1}(d^L f_1(b)) b , d^Rf_2(b)\right)_\D,
\qquad
\forall f_1, f_2\in C^\infty(B),\, \forall b\in B.
\label{2.17}\ee
This is obtained from (\ref{2.5}) and (\ref{2.6}) using that  if  $\Phi = \Lambda_R^* f$ for
some $f\in C^\infty(B)$, then
\be
\nabla^L \Phi(K)=-g_L \left(d^R f(b_R)\right) g_L^{-1},
\qquad
\nabla^R \Phi(K)=-b_R \left(d^R f(b_R)\right) b_R^{-1}.
\label{2.18}\ee
The induced Poisson-Lie structure on $G$ permits the analogous formula
\be
\{ \phi_1, \phi_2\}_G(g)=  \left(  g^{-1}(d^L \phi_1(g)) g, d^R\phi_2(g)\right)_\D,
\qquad
\forall \phi_1, \phi_2\in C^\infty(G),\, \forall g\in G,
\label{2.19}\ee
which can be conveniently rewritten as
\be
\{ \phi_1, \phi_2\}_G(g)=  \langle \bD^R \phi_1(g), R^\ri(\bD^R\phi_2(g))\rangle_\G
-\langle \bD^L \phi_1(g), R^\ri(\bD^L\phi_2(g))\rangle_\G
\label{2.20}\ee
by virtue of (\ref{2.15}).
In the last formula only $G$ features explicitly, while (\ref{2.17}) and (\ref{2.19})
are formulated relying on conjugations defined in the group $D$.

The Poisson bracket $\{.,.\}_+$ closes also on $\Lambda_L^* C^\infty(B)$ and on
$\Xi_L^* C^\infty(G)$.
In fact, by using $\Lambda_L$ and $\Xi_L$ (\ref{Iwasawa})  one obtains the same Poisson-Lie structures
on $B$ and on $G$ as by means of (2.16).
Another important fact is that the elements of $\Lambda_R^* C^\infty(B)$ (respectively
$\Xi_R^* C^\infty(G)$) commute with the elements of
$\Lambda_L^* C^\infty(B)$  (respectively $\Xi_L^* C^\infty(G)$) with respect to $\{.,.\}_+$.

Finally, we recall from \cite{K} the so-called
quasi-adjoint
 Poisson action of $G$ on $(D,\omega_+)$.
 The corresponding  action map $G\times D\to D$ sends $(g,K)$ to $g\triangleright K$
 defined by
\be
g \triangleright K :=gK\Xi_R(g\Lambda_L(K)),
\quad \forall g\in G,\, \forall K\in D.
\label{2.21}\ee
The composition property,
 $g_1 \triangleright (g_2\triangleright K) = (g_1 g_2) \triangleright K$
 for all $g_1, g_2 \in G$ and $K\in D$, can be checked by using (\ref{2.6}) and (\ref{Iwasawa}).
 The $G$-action $\triangleright$  admits the equivariant Poisson-Lie moment map
 $\Lambda:D\to B$ given by
 \be
 \Lambda (K)=\Lambda_L(K)\Lambda_R(K), \qquad \forall K\in D.
 \label{2.22} \ee
This means that $\Lambda$  enjoys the following two properties.
First,
\be
X_D[\Phi]=(X,\{\Phi,\Lambda \}_{+}\Lambda^{-1})_\D,\quad
\forall X\in\G,\,  \forall \Phi\in C^\infty(D),
\label{2.23}\ee
where $(X_D[\Phi])(K):={\frac{d}{ds}}\Phi(e^{sX}\trl K)\vert_{s=0}$ for all $K\in D$.
Second,
\be
\Lambda(g\trl K)=\Dress_g(\Lambda(K)),
\quad \forall g\in G, \, \forall K\in D,
\label{2.24}\ee
where we use the dressing action of $G$ on $B$ that operates as
\be
\Dress_g(b):=\Lambda_L(gb), \quad \forall g\in G,\, \forall b\in B.
\label{2.25}\ee
The  quasi-adjoint action  (\ref{2.21}) was found in \cite{K}
by the following method.
One first observes that the map $\Lambda$
defined by (\ref{2.22}) satisfies (\ref{2.24}) and  it
generates via (\ref{2.23})  an  infinitesimal action of $\G$ on $D$.
These statements are implied \cite{Lu} by the fact that $\Lambda: D \to B$ is a
Poisson map, which is obvious.
The resulting infinitesimal action was then integrated, and
the $G$-action obtained in this way automatically has the Poisson property, i.e.,
the action map $G\times D\to D$ is a Poisson map.

\medskip
\noindent
{\bf Remark 2.1.}
It follows from  (\ref{2.21})  that, like for the ordinary adjoint action,
the central $U(1)$ subgroup of $G=U(n)$ acts trivially, and the
factor group $U(n)/U(1)$ acts effectively.
It is also easy to check (see Appendix A) that
the map $\Lambda$ (\ref{2.22})   takes its values in the Poisson-Lie
subgroup
$SB$ of $B$ consisting of the elements of determinant one.
Consider the projection
\be
\pi_D: D \to \bar D:= D/GL(1,\bC),
\label{Rem1}\ee
where  $GL(1,\bC)$ denotes the center of $D=GL(n,\bC)$,
and define subgroups of $\bar D$ by
\be
\bar G:= \pi_D(G)
\quad\hbox{and}\quad
\bar B:= \pi_D(B) = \pi_D(SB).
\label{Rem2}\ee
The groups $\bar G$ and $\bar B$ sit in $\bar D$ in quite the same
way as $G$ and $B$ sit in $D$, and therefore are dual to each
other in the Poisson-Lie sense. In fact, $\bar D$
carries a symplectic structure
inherited from $(D, \omega_+)$, whereby it can be regarded as
the Heisenberg double constituted by $\bar G$ and $\bar B$.
The projection $\pi_D$ gives rise to natural isomorphisms
\be
U(n)/U(1) \simeq \bar G
\qquad
\hbox{and}\qquad  SB \simeq \bar B.
\label{Rem3}\ee
By using these identifications, the map $\Lambda: D \to SB$ given by (\ref{2.22})
yields the Poisson-Lie moment map for the
quasi-adjoint action of $U(n)/U(1)$ on $D$.

\subsection{Integration algorithms for the free flows}

Our principal aim now is to present the flows of two families of commuting Hamiltonians
on the Heisenberg double that are invariant with respect to the quasi-adjoint
action of $G$.

The Hamiltonians of our interest
form the rings $\H$ and $\hat\H$ defined by
\be
\H:=\Lambda_R^*C^\infty(B)^c
\qquad\hbox{and}\qquad
\hat\H:=\Xi_R^*C^\infty(G)^G.
\label{2.26}\ee
Here  $C^\infty(B)^c$ denotes the  center of
the Poisson-Lie structure on $C^\infty(B)$ and
$C^\infty(G)^G$ contains the functions on $G$ that are invariant with respect to
the standard adjoint action of $G$ on $G$.
It is obvious that all elements of $\H$ mutually Poisson commute
which each other. The Poisson commutativity  of $\hat \H$ also takes places,
because
$\bD^L \phi = \bD^R \phi$ for any $\phi\in C^\infty(G)^G$, and
hence  (\ref{2.20}) implies that
$\{ \phi_1, \phi_2\}_G =0$ for any
$\phi_1, \phi_2 \in C^\infty(G)^G$.
Note, however, that the elements of  $C^\infty(G)^G$ do not lie in   the center of the
Poisson-Lie structure on $C^\infty(G)$ in general.

It is well known and is readily seen from (\ref{2.17}) and (\ref{2.25}) that
\be
C^\infty(B)^c = C^\infty(B)^G
\label{2.28}\ee
with   the dressing-invariant functions on the right-hand side.
Moreover, if $\mathrm{inv}: B\to B$ is the (anti-Poisson) inversion map, then
one can verify that $\mathrm{inv}^*$ stabilizes $C^\infty(B)^c$ and
\be
\Lambda_R^* f = \Lambda_L^* (f\circ \mathrm{inv}),
\qquad
\forall f\in C^\infty(B)^c.
\label{2.29}\ee
In particular, it follows that
\be
\H:=\Lambda_R^*C^\infty(B)^c =\Lambda_L^*C^\infty(B)^c.
\label{2.30}\ee
To obtain another useful characterization of $\H$, consider the diffeomorphism, $\mathfrak{P}$, from $B$ to
the space of
Hermitian positive definite matrices, ${\cal P}$, defined by
\be
\mathfrak{P}(b):=bb^\dagger,  \qquad \forall b\in B.
\label{2.31}\ee
 The map $\mathfrak{P}$ intertwines the dressing action (\ref{2.25})  on $B$
with the ordinary conjugation action on ${\cal P}$,
\be
\mathfrak{P}(\Dress_gb)=g\mathfrak{P}(b)g^{-1},
\qquad \forall g\in G,\, \forall b\in B,
\label{2.32}\ee
whereby the elements of $C^\infty(B)^c$ correspond to the spectral-invariants
$C^\infty({\cal P})^G$ on ${\cal P}$.

It is of crucial importance that all Hamiltonians in
$\H$ and in $\hat\H$ are
 invariant under the quasi-adjoint action (\ref{2.21}) on the double. This holds  since
for all $g\in G$ and  $K\in D$ one has
\be
\Lambda_L (g \triangleright K)=\Dress_g(\Lambda_L(K)),
\qquad
\Xi_R (g \triangleright K)= \Xi_R(g \Lambda_L(K))^{-1} \Xi_R(K) \Xi_R(g \Lambda_L(K)).
\label{2.33}\ee
The first relation and (\ref{2.28}), (\ref{2.30}) imply the $G$ invariance of the elements of $\H$,
which also follows directly from
$\Lambda_R (g \triangleright K)=\Dress_{\Xi_R(g \Lambda_L(K))^{-1}}(\Lambda_R(K))$.

Now we are ready to show that the evolution equations of the
Hamiltonian systems $(D,\omega_+,H)$ and $(D,\omega_+,\hat H)$ can be explicitly solved for any $H\in \H$ and
any $\hat H\in \hat\H$.
We start with $\H$.

\medskip
\noindent
{\bf Proposition 2.1.}\emph{
The flow induced on the Heisenberg double $D$ by the Hamiltonian
$H = \Lambda_R^* f$ with an arbitrary $f\in C^\infty(B)^c$
is given by
\be
K(t)= g_L(0) \exp\left[-t d^R f(b_R(0)) \right] b^{-1}_R(0).
\label{2.35}\ee
In terms of the Iwasawa decompositions (\ref{2.6}), the flow can be written as
\be
g_L(t) = g_L(0)\exp\left[-t d^R f(b_R(0)) \right],
\qquad
b_R(t)=b_R(0),
\label{2.36}\ee
\be
g_R(t) = \exp\left[-t d^L f(b_L(0)) \right] g_R(0),
\qquad
b_L(t)=b_L(0).
\label{2.37}\ee}

\medskip
\noindent
{\bf Proof.}
Equation (\ref{2.5}) entails that the Hamiltonian vector field,
$V_\Phi$, generated by an arbitrary $\Phi\in C^\infty(D)$ has the form
\be
V_\Phi(K)= K \rho(\nabla^R \Phi(K)) + \rho(\nabla^L\Phi(K)) K.
\label{2.38}\ee
Notice  from (2.17) that
\be
\pi_\B\left(b \left(d^R f(b)\right) b^{-1}\right)=0,
\qquad
\forall b\in B,\, \forall f\in C^\infty(B)^c.
\label{2.39}\ee
By combining these formulae and (\ref{2.18}), we obtain
\be
V_\Phi(K) =-g_L \left( d^R f(b_R)\right) b_R^{-1},
\quad\hbox{if}\quad
\Phi=\Lambda_R^* f,\, f\in C^\infty(B)^c.
\label{2.40}\ee
At the same time, since $\Lambda_R: D\to B$ is a Poisson map, we also have $V_\Phi(b_R)=0$.
It follows immediately  that the flow is given by (\ref{2.35}),
which obviously translates into (\ref{2.36}).
The alternative formula (\ref{2.37}) can be established using that
$b_L(t)=b_L(0)$ by (\ref{2.30}), and that
 $d^L f(b) = b (d^R f(b)) b^{-1}$ for any $f\in C^\infty(B)^c$.
\emph{Q.E.D.}
\medskip

Proposition 2.1 says
that the Poisson-Lie momenta $b_L$, $b_R$ are constants of motion and both
`position-like' variables $g_L$, $g_R$
follow Killing geodesics on $G$.
In a special case, this statement first appeared in \cite{Zakr}.

Now we turn to  the flows associated with the conjugation-invariant functions
on $G$.  We begin by treating this problem on the Poisson-Lie group $G$.

\medskip
\noindent
\textbf {Lemma 2.1.}
\emph{For $\phi\in C^\infty(G)^G$,
consider the Hamiltonian evolution equation on $(G, \{.,.\}_G)$,
\be
\dot{g} := \{ g, \phi\}_G = [ g, R^\ri(\bD\phi(g))],
\label{2.41}\ee
where we denote
$\bD^L\phi = \bD^R\phi$ simply by $\bD\phi$.
Taking an arbitrary initial value $g(0)$, let the curves $\beta(t)\in B$ and $\gamma(t)\in G$
be the (unique, smooth) solutions of the factorization problem
\be
e^{\ri t \bD\phi(g(0))} = \beta(t) \gamma(t).
\label{2.42}\ee
Then the solution of (\ref{2.41}) with the initial value $g(0)$ is given by
\be
g(t)= \gamma(t) g(0) \gamma(t)^{-1}.
\label{2.43}\ee}

\medskip\noindent
{\bf Proof.} By taking the derivative of (\ref{2.43}), we obtain
\be
\dot{g}(t)= [ \dot\gamma(t) \gamma(t)^{-1}, g(t)].
\label{2.44}\ee
On the other hand, by taking the derivative of (\ref{2.42}) we obtain
\be
\beta(t) \gamma(t) (\ri \bD\phi(g(0))) =  \dot \beta(t)\gamma(t) + \beta(t)\dot\gamma(t).
\label{2.45}\ee
This implies
\be
\ri \gamma(t)\bD\phi(g(0)) \gamma(t)^{-1} = \ri \bD\phi(g(t)) = \beta(t)^{-1} \dot \beta(t)
+ \dot\gamma(t) \gamma(t)^{-1},
\label{2.46}\ee
where the first equality follows from the invariance property of $\phi$.
We  see from (\ref{2.46}) that
\be
\dot\gamma(t) \gamma(t)^{-1} = \pi_\G(\ri \bD\phi(g(t))) = - R^\ri(\bD\phi(g(t)),
\label{2.47}\ee
which concludes the proof. \emph{Q.E.D.}

\medskip
\noindent
\textbf {Proposition  2.2.} \emph{The flow $K(t) = b_L(t) g_R(t)^{-1}$ induced on the Heisenberg double
by the Hamiltonian
$\hat H = \Xi_R^* \phi$ with an arbitrary  $\phi\in C^\infty(G)^G$
is given by
\be
g_R(t) = \gamma(t) g_R(0) \gamma(t)^{-1},
\qquad
b_L(t) = b_L(0) \beta(t)
\label{2.48}\ee
in terms of the solutions $\gamma(t)\in G$, $\beta(t) \in B$ of the factorization problem
$e^{\ri t \bD\phi(g_R(0))} = \beta(t) \gamma(t)$.}

\medskip
\noindent
{\bf Proof.} Consider first an arbitrary `collective Hamiltonian' on $D$ of the
type
\be
\Phi = \Xi_R^* \phi,
\qquad
\phi\in C^\infty(G).
\label{2.49}\ee
In this case one obtains directly from the definitions that
\be
\nabla^L \Phi(K)= - b_L (d^R \phi(g_R)) b_L^{-1},
\qquad
\nabla^R \Phi(K)= - g_R (d^R\phi(g_R)) g_R^{-1}.
\label{2.50}\ee
With the aid of these relations,
the evolution equation $\dot K = V_\Phi(K)$ (\ref{2.38})  can be
spelled out as
\bea
&& \dot g_R = g_R R^\ri(\bD^R \phi(g_R)) - R^\ri(\bD^L\phi(g_R)) g_R,
\label{2.51}\\
&& \dot b_L = b_L (d^R \phi(g_R))=b_L \pi_\B (\ri \bD^R \phi(g_R)).
\label{2.52}\eea
Besides (\ref{2.50}), we also used (\ref{2.13}) and (\ref{2.14}) to get this system of equations.
Notice that (\ref{2.51}) is just the Hamiltonian evolution equation generated by $\phi$ on $(G, \{.,.\}_G)$.
If we now  assume that $\phi\in C^\infty(G)^G$, then the desired solution of the last system of equations
is easily found with the aid of (\ref{2.43})
and (\ref{2.46}). This yields the flow as claimed in (\ref{2.48}).
\emph{Q.E.D.}

\medskip
\noindent
\textbf {Corollary 2.1.} \emph{The integral curve of the Hamiltonian
$\hat H= \Xi_R^* \phi$, $\phi\in C^\infty(G)^G$  satisfies
\be
K(t) K^\dagger(t) = b_L(t) b_L(t)^\dagger = b_L(0) e^{2 \ri t \bD\phi(g_R(0))}b_L(0)^\dagger.
\label{2.53}\ee
}
\medskip
\noindent
{\bf Proof.}
Combine (\ref{2.48}) with
$b_L(0) \beta(t) (b_L(0) \beta(t))^\dagger = b_L(0) \beta(t) \gamma(t) ( b_L(0) \beta(t) \gamma(t))^\dagger$.
 \emph{Q.E.D.}

\medskip

We finish this section with a few comments.
First we notice from the statement below (\ref{2.32}) that $C^\infty(B)^c$
is functionally generated by the (not all independent) invariants
\be
f_k(b) := \frac{1}{2k} \tr (b b^\dagger)^k,
\qquad
\forall k\in \bZ^*.
\label{2.54}\ee
It is also clear that $C^\infty(G)^G$ is generated by the functions
\be
 \phi_k(g):=\frac{1}{2k} \tr(g^k+g^{-k}),\quad
  \phi_{-k}(g):=\frac{1}{2k \ri} \tr(g^k - g^{-k}),
\quad \forall k\in \bZ_+.
\label{2.55}\ee
The flows of
the generators $H_k:= \Lambda_R^* f_k$ of $\H$ and
$\hat H_k:= \Xi_R^* \phi_k$ of $\hat \H$
can be written down explicitly
using that
\be
d^R f_k(b) = \ri (b^\dagger b)^k,
\quad
d^L f_k(b) = \ri (b b^\dagger)^k,
\quad \forall k\in \bZ^*,
\label{2.56}\ee
\be
\bD \phi_k(g) =\frac{1}{2}( g^k - g^{-k}),
\quad
\bD \phi_{-k}(g) =\frac{1}{2\ri}( g^k + g^{-k}),
\quad
\forall k\in \bZ_+.
\label{2.57}\ee
For later reference, we record that if one considers a real linear combination
$\phi := \sum_{k\neq 0} \mu_k \phi_k$, and writes it with some analytic function $\chi$ in the form
$\phi(g) = \tr ( \chi(g)) + \mathrm{c.c.}$, then one has
\be
\bD \phi (g) = g \chi'(g) - (g \chi'(g))^\dagger.
\label{2.58}\ee

We note from the above that the
Hamiltonians $H_k$ and $\hat H_k$, and more generally the elements of
the rings $\H$  and $\hat\H$ (\ref{2.26}) that they generate,
are spectral-invariants
of the respective matrix functions ${\cal L}$ and $\hat {\cal L}$  on
the Heisenberg double $D$ defined by
\be
{\cal L}: K\mapsto  b_R b_R^\dagger
\qquad\hbox{and}\qquad
\hat {\cal L}: K \mapsto g_R\label{Unlax}
\ee
with the Iwasawa decompositions in (\ref{2.6}).
For our purpose, it will be  fruitful to view  ${\cal L}$ and $\hat {\cal L}$   as
`unreduced Lax matrices' and
the following convention  will also prove to be very convenient.

 \medskip
\noindent {\bf Definition 2.1.} \emph{By using the previous notations
(\ref{2.26})  and (\ref{Unlax}),
we define the collections of Hamiltonian systems
\bea
&&(D, \omega_+, \cL):= \{\, (D, \omega_+, H) \, \vert\, H\in \H\,\},\nonumber\\
&&(D, \omega_+, \hat \cL):= \{\, (D, \omega_+, \hat H) \, \vert\, \hat H\in \hat \H\,\},
\label{candef}\eea
and henceforth refer to these collections  as
the `canonical free systems'.}

\medskip

Any member of the above collections is integrable in the obvious sense
that one can directly write down its Hamiltonian flow, as given by Propositions 2.1 and 2.2.
In the rest of the paper we shall study the symplectic reduction
of the canonical free systems, i.e., the simultaneous reduction
of all the Hamiltonian systems that constitute them.
The main advantage of our, somewhat non-standard,  notation (\ref{candef})  is that
it suggests that  one should directly reduce
the Lax matrices $\cL$ and $\hat \cL$, instead of separately reducing
the Hamiltonians that they generate.
In this respect,
observe from (\ref{2.33}) and the sentence afterwards
that the quasi-adjoint action (\ref{2.21}) operates by similarity transformations
 on  ${\cal L}$ and on $\hat {\cal L}$ in (\ref{Unlax}).
Since, as generators of commuting Hamiltonians,
 any two Lax matrices that are related by a similarity transformation are equivalent,
one can indeed use the quasi-adjoint
 action to reduce the (equivalence classes of the) Lax matrices $\cal L$ and $\hat {\cal L}$.
 The usefulness of this point of view will become clear in Section 3.

Finally, for clarity, let us remark that the elements of the Abelian subalgebra
$\Xi_L^* C^\infty(G)^G$
of the Poisson algebra $C^\infty(D)$ are in general not invariant under
the quasi-adjoint action (\ref{2.21}). Indeed
\be
\quad
\Xi_L (g \triangleright K)= g \Xi_L(K) \Xi_L(\Lambda_R(K)^{-1} \Xi_R(g \Lambda_L(K)))
\label{2end}\ee
is not conjugate to $\Xi_L(K)$ in general.
However,
the elements of $\Xi_L^* C^\infty(G)^G$
are actually invariant with respect
to the alternative quasi-adjoint action \cite{K} of $G$ on $D$ that  can be associated
with the `flipped moment map'
$\Lambda':= \Lambda_R \Lambda_L: D \to B$.
(The elements of $\H$ (\ref{2.30}) are invariant under both
quasi-adjoint actions.)
Since $\Lambda= \Lambda_L \Lambda_R$ can be converted into $\Lambda'$
by  inversion on the group $D$,
it is sufficient to consider only the quasi-adjoint
action given by (\ref{2.21}).

\section{Reduction of the canonical free systems}
\setcounter{equation}{0}

Our goal here is to present the results that permit the identification of a certain symplectic
reduction of the canonical free systems
$(D, \omega_+, \cL)$ and
 $(D, \omega_+, \hat \cL)$ of Definition 2.1
  with the Ruijsenaars-Schneider system
 $(P, \omega, L)$ (\ref{1.1}) and with a natural extension of the dual system
 $(\hat P, \hat \omega, \hat L)$ (\ref{1.3}), respectively.
 For this purpose we must take $D=GL(n,\bC)$, but it is worth stressing that
 the preliminaries described in Section 2 remain valid in a more general context.

We first review the necessary  theoretical  background  concerning symplectic reduction based on Poisson-Lie
symmetry   with an equivariant moment map \cite{Lu}.
(The notations $\barG$, $\bar B$, $\Lambda$ used in this overview anticipate
the application studied later on.)
Consider a symplectic manifold $M$ acted upon  smoothly and {\it effectively}  by
a {\it compact}   Poisson-Lie
group
$\barG$  in such a way that the  action map
$\Psi: \barG\times M \to M$
is Poisson and
  choose a regular value $\nu\in \bar B$ in the image of the
moment
map\footnote{The moment map is a Poisson map into the Poisson-Lie group $\bar B$ dual to $\bar G$.
It generates the $\bar G$ action via the Poisson bracket on $M$ by
$X_M[f] = \langle X, \{f, \Lambda\}_M \Lambda^{-1} \rangle$, where $X_M$ is the vector field on $M$
corresponding to $X\in \mathrm{Lie}(\barG)$, $f\in C^\infty(M)$, and $\langle .,.\rangle$
is the dual pairing.}
$\Lambda:M\to \bar B$.
The  $\Lambda$-preimage $F_\nu$ of the point $\nu$ is then  an embedded  submanifold of $M$. The maximal
subgroup $\barG_\nu < \barG$ which leaves $F_\nu$ invariant  is called the gauge group.
If $\barG_\nu$ acts freely on
$F_\nu$ then, as  is well known, there exists a unique manifold structure on
the space of orbits, $F_\nu/\barG_\nu$, such that the canonical projection
\be
\pi: F_\nu \to F_\nu/\barG_\nu
\label{B1}\ee
is a smooth submersion, i.e.,
$F_\nu(F_\nu/\barG_\nu, \barG_\nu, \pi)$ is a principal fiber bundle.
 Let us note that
at every point $m\in F_\nu$ the tangent space $T_mF_\nu$ has
 the vertical subspace
\be
V_m F_\nu \subset T_m F_\nu
\label{B2}\ee
generated by the infinitesimal action of $\barG_\nu$.
Smooth functions on $F_\nu/\barG_\nu$
correspond to smooth  $\barG_\nu$-invariant functions on $F_\nu$.
The manifold structure on $F_\nu/\barG_\nu$ is constructed  with the aid of local
cross sections for the $\barG_\nu$ action \cite{DK}.

In fact, it turns out that  $F_\nu/\barG_\nu$ is  a  symplectic manifold.  It is referred to as the
reduced symplectic manifold (or reduced phase space) and the symplectic form $\Omega_\nu$
on  it is uniquely determined by the requirement
\be
\Omega\vert_{F_\nu}=\pi^*\Omega_\nu.
\label{sr}\ee
 Here $\Omega\vert_{F_{\nu}}$ denotes the pull-back of the
 original symplectic form $\Omega$ of $M$ on the submanifold
 $F_\nu \subset M$.

The Hamiltonian flow induced by any $\barG$-invariant function
$H \in C^\infty(M)^{\barG}$ preserves $F_\nu$, and
the restricted flow is projectable from $F_\nu$ to $F_\nu/\barG_\nu$.
The projection gives the flow associated with the reduced Hamiltonian
$H^\nu\in C^\infty(F_\nu/\bar G_\nu)$ (for which  $H\vert_{F_\nu} = H^\nu \circ \pi$)
by means of $\Omega_\nu$.
It further follows that the \emph{involutivity} of a set of $\barG$-invariant Hamiltonians
$\{ H_j\} \subset C^\infty(M)^\barG$ is inherited by the corresponding set
of reduced Hamiltonians $\{ H_j^\nu\} \subset C^\infty(F_\nu/\barG_\nu)$.

\medskip

In concrete examples  one wishes to exhibit models of the
reduced symplectic manifold $(F_\nu/\barG_\nu,\Omega_\nu)$.
In principle, any symplectic manifold that is globally symplectomorphic to
$(F_\nu/\barG_\nu,\Omega_\nu)$ can serve as a model, but the real aim is to
construct models as explicitly as possible.
The simplest situation occurs when the manifold $F_\nu$  is
a {\it trivial}  $\barG_\nu$-bundle.  Then the reduced
symplectic manifold $(F_\nu/\barG_\nu,\Omega_\nu)$    can be modelled by
any global cross section of the $\barG_\nu$ action on  $F_\nu$.
To be more precise, we present the following  (standard) result.

\medskip

\noindent {\bf Lemma 3.1.}   \emph{Suppose that $(\bar P,\bar \om)$ is a symplectic manifold
 and $\cJ:\bar P\to F_\nu$  is a smooth injective map such that
  \begin{enumerate}
 \item{ $\cJ^*(\Omega\vert_{F_\nu})=\bar \om$, }
 \item{the image $S:= \{ \cJ(y)\,\vert\, y\in \bar P\,\}$
intersects every $\barG_\nu$-orbit in $F_\nu$ exactly in one point.}
\end{enumerate}
Then the map $ \pi \circ \cJ: \bar P \to F_\nu/\barG_\nu$ is a symplectic diffeomorphism,
and
$(\bar P,\bar \om)$ can thus serve as a model of the reduced phase space $(F_\nu/\barG_\nu,\Omega_\nu)$.}

\medskip

\noindent {\bf Proof.} The procedure of symplectic reduction rests on the fact that
\be
\mathrm{Ker}_m(\Omega\vert_{F_\nu})
= V_m F_\nu,
\qquad
\forall m\in F_\nu,
\label{B4}\ee
where $\mathrm{Ker}_m(\Omega\vert_{F_\nu})$ is the annihilator of
$\Omega\vert_{F_\nu}$ in $T_m F_\nu$.
This equation and $\cJ^*(\Omega\vert_{F_\nu})=\bar\om$  imply
that the tangent (derivative) map
$T_y \cJ\equiv (D\cJ)(y): T_y \bar P \to T_{\cJ(y)} F_\nu$
is injective and
\be
V_{\cJ(y)} F_\nu \cap T_y \cJ(T_y \bar P)=\{ 0\}, \qquad \forall y\in \bar P.
\label{B5}\ee
Indeed, the hypothesis that, at some $y\in \bar P$,
  $T_y \cJ(Y)  \in V_{\cJ(y)}F_\nu$
for a non-zero $Y\in T_y \bar P$  would entail that
$\cJ^*( \Omega\vert_{F_\nu})(Y, Z)  = 0$ for all $Z\in T_y \bar P$.
However, this is excluded by the non-degeneracy of  $\cJ^*( \Omega\vert_{F_\nu})$.
The injectivity of $T_y \cJ$ means that the map $\cJ: \bar P \to F_\nu$ is
an immersion.
Next, being the composition of two smooth maps, $\pi \circ \cJ: \bar P \to F_\nu/\barG_\nu$ is smooth
and we see from (\ref{B5}) that
it is also an immersion.
Therefore,
since $\pi \circ \cJ$ is  injective and surjective by assumption,
it must be a submersion, which in particular requires that
\be
T_{\cJ(y)} F_\nu = V_{\cJ(y)} F_\nu \oplus T_y \cJ(T_y \bar P),\qquad \forall y\in \bar P.
\label{B3}\ee
It is well known that the one-to-one smooth submersions are
precisely the diffeomorphisms.   Hence we conclude that the map $\pi \circ \cJ$ is a diffeomorphism.
Finally,
using  (\ref{sr}), we obtain
\be
(\pi\circ\cJ)^*\Omega_\nu=\cJ^*(\pi^*\Omega_\nu)=\cJ^*(\Omega\vert_{F_\nu})=\bar \om.
\ee
\emph{Q.E.D.}

 \medskip

\noindent {\bf Remark 3.1.}
 On account of (\ref{B3}), the smooth
one-to-one map $\Psi \circ (\mathrm{id}_{\barG_\nu}, \cJ):\barG_\nu \times \bar P \to F_\nu$ is
  again a submersion. This implies  that  the map
$\Psi \circ (\mathrm{id}_{\barG_\nu}, \cJ)$
is a diffeomorphism. Hence the map $\cJ$ is an embedding and
the restriction of the  projection $\pi$ to $S=\cJ(\bar P)$
yields a diffeomorphism, $\pi_S: S \to F_\nu/\barG_\nu$. Moreover,
if we define $\sigma: F_\nu/\barG_\nu \to F_\nu$  by
$\sigma(b) := \cJ(y)$
for  the unique $y\in \bar P$ such that
$b = \pi(\cJ(y))$, then $\sigma$
is a smooth global section of the bundle $\pi: F_\nu \to F_\nu/\barG_\nu$ in the
usual sense, i.e., $\sigma\in C^\infty(F_\nu/\barG_\nu, F_\nu)$ and
$\pi \circ \sigma = \mathrm{id}_{F_\nu/\barG_\nu}$.
 This follows  by noting that $\sigma = \iota_S \circ \pi_S^{-1}$,
where $\iota_S: S \to F_\nu$ is the tautological injection.

 \medskip

\noindent {\bf Remark 3.2.}
We call the map $\cJ$  of Lemma 3.1 a \emph{global cross section}  of the $\barG_\nu$
  action on $F_\nu$.
   The same term can be used to refer to the image $S$ of $\cJ$, too,
    but here we adopt the more widespread
   terminology that refers to $S$ as a \emph{global gauge slice}.
   Note that $(S, \Omega\vert_S)$ can be also thought of as a model of the reduced phase space,
   since it is symplectomorphic to $(\bar P, \bar \omega)$ by $\cJ$.

\medskip

 From now on we take the  unreduced symplectic manifold to be
the Heisenberg double,
\be
 (M,\Omega) := (\gln,\om_+),
\label{identify1}\ee
and identify the  effectively acting symmetry group $\barG$    with  $U(n)$ divided by its center.
As explained in Remark 2.1, $\bar G$ acts by  the quasi-adjoint action (\ref{2.21}) according to
\be
 \Psi([g],K):= g \triangleright K,
 \qquad  \forall [g]\in \bar G, \,\forall K\in GL(n,\bC),
\label{identify2} \ee
where $g\in U(n)$ is an arbitrary representative of $[g]\in U(n)/U(1)$.
In Definition 2.1 of Section 2,  we   introduced   two families of $\barG$ Poisson-Lie symmetric
 Hamiltonian systems, namely
$(\gln,\om_+,{\cal L})$ and $(\gln,\om_+,\hat{\cal L})$.
Since these  canonical free systems  have  the same symplectic structure $\omega_+$
and they are Poisson-Lie symmetric
with respect to   the same action of $\bar G$,
we can consider their simultaneous symplectic
 reduction   based on  a moment map value $\nu$.  In our letter
 \cite{LMP}, we   identified
 the (parameter-dependent)  value $\nu(x)\in  \bar B$   such that the symplectic
 reduction of   the  system $(\gln,\om_+,{\cal L})$  gives   the   original
 Ruijsenaars-Schneider system $(P, \omega, L)$  (\ref{1.1}).
 An enhanced formulation of this result is  given
  by Theorems 3.1 and 3.2 below.
  Later on, we  shall formulate
  Theorems 3.3 and 3.4, which claim that  the  symplectic
 reduction of    $(\gln,\om_+,\hat{\cal L})$  with respect to the same $\nu(x)$  gives an
 extension of the dual Ruijsenaars-Schneider system $(\hat P, \hat \omega, \hat L)$  (\ref{1.3}).

 In order to formulate  precisely the above mentioned theorems,   first we have
 to recall
 some notations and results from  our previous paper
\cite{LMP}.  Thus let $x$ be a real non-zero parameter and denote by $\nu(x)$ the
element of the group $\barB \simeq SB < B $ (remember (\ref{Rem3})) such that
 \be
 \nu(x)_{kk}=1, \quad \forall k, \qquad \nu(x)_{kl}= (1-e^{-x})e^{\frac{(l-k)x}{2}},
 \quad \forall k<l.\label{bx}\ee
There holds the equation
 \be \nu(x)\nu(x)^\dagger=e^{-x}\left[ \1_n+ \frac{e^{nx}-1}{n}v(x)  v(x)^\dagger \right],
\label{3.13}\ee
 where the components of the real column vector $v(x)$ read
  \be
  v_k(x)=\sqrt{\frac{n(e^x-1)}{1-e^{-nx}}}e^{-{\frac{kx}{2}}}\,,
  \qquad \forall k=1,..., n.
  \label{vx}\ee
The isotropy subgroup $G_{\nu(x)} < U(n)$ is the direct product
$G_{v(x)} \times U(1)$, where $G_{v(x)}< U(n)$ is the isotropy group of  $v(x)\in \bC^n$
and $U(1)$ is the center of $U(n)$.
The corresponding subgroup $\bar G_{\nu(x)}$ of the effective symmetry group   $ \bar G = U(n)/U(1)$
admits the natural isomorphisms
\be
\barG_{\nu(x)} \simeq G_{\nu(x)}/ U(1) \simeq G_{v(x)},
\label{Gv}\ee
and therefore we can identify  $\barG_{\nu(x)}$ with $G_{v(x)}$.

\medskip
Let $\Lambda:GL(n,\bC)\to  \bar B\simeq SB$ (\ref{Rem3})  be the moment map    of the
 quasi-adjoint action (\ref{identify2})  of $\barG=U(n)/U(1)$.
 With $\nu(x)$  in (\ref{bx}),
 consider the symplectic reduction of the Heisenberg double
 $(GL(n,\bC),\omega_+)$  defined by imposing the constraint
 \be \Lambda(K):=\Lambda_L(K)\Lambda_R(K)=\nu(x),\quad K\in GL(n,\bC). \label{MMC}\ee
The  associated  gauge group  is given by $\barG_{\nu(x)}$ (\ref{Gv}), and we have the following basic lemma.

 \medskip

 \noindent  {\bf Lemma 3.2 (\cite{LMP}).} {\it  The set $F_{\nu(x)}$ consisting
 of the solutions of the moment map
 constraint (\ref{MMC}) is an
 embedded submanifold of $\gln$ and the  compact  group $\bar G_{\nu(x)}$
 acts freely on it.}

 \medskip

 Lemma 3.2  was proven in \cite{LMP} by  finding explicitly all solutions
 of the moment map constraint
 (\ref{MMC}). For completeness,   in Appendix A
we offer an alternative proof of the fact that $F_{\nu(x)}$ is an embedded submanifold  by showing that  the
element $\nu(x)$  is a \emph{regular value} of the moment map $\Lambda$.
It follows from the lemma that the orbit space $F_{\nu(x)}/\bar G_{\nu(x)}$ is a smooth manifold:
the base of the principal fiber bundle with total space $F_{\nu(x)}$ and structure group
$\barG_{\nu(x)}$. For any $H\in \cal H$ (\ref{2.26}) (respectively $\hat H\in\hat \H$),
the reduction of the Hamiltonian system
$(\gln,\om_+,H)$    yields a system
with \emph{complete} Hamiltonian flow on the reduced phase space, which can be
constructed by  projecting the original flow given by Propositions 2.1
(respectively Proposition 2.2).
Before giving a characterization of  the reduction of the canonical free system
$(\gln,\om_+,\cL)$ of
Definition 2.1, next we present a convenient description  of the phase space $(P,\omega)$
of the original Ruijsenaars-Schneider system (\ref{1.1}).

Let $\bT_n^0$ denote the regular part of the standard maximal torus $\bT_n < G=U(n)$.
The symmetric group $S(n)$ acts freely on $\bT_n^0$, any $\sigma\in S(n)$ acts by
permuting the diagonal entries  of the elements
$T=\diag(T_1,\ldots, T_n)\in \bT_n^0$.
Thus the corresponding space of $S(n)$-orbits
\be
Q(n):= \bT^0_n/ S(n)
\label{Qn}\ee
is a smooth manifold. This is the set of unordered $n$-tuples of pairwise distinct points
of the circle $S^1= U(1)$.
The phase space of the trigonometric Ruijsenaars-Schneider  system, when interpreted as a many-body system
of indistinguishable particles, is in fact given by
\be
(P,\omega):= (T^* Q(n), \Omega_{T^* Q(n)}),
\label{3.16}\ee
where $\Omega_{T^*Q(n)}$ is the canonical symplectic form of the cotangent bundle $T^* Q(n)$.
Note that $S(n)$ acts freely
on $T^* \bT_n^0$, too, by the cotangent lift of the $S(n)$-action on $\bT_n^0$.
If we use the realization
\be
T^* \bT_n^0 \simeq \bT_n^0 \times  \bR^n = \{ ( e^{2\ri q}, p)\},
\label{3.17}\ee
then this $S(n)$-action operates by the simultaneous permutations of the
entries of $e^{2\ri q}$ and $p$. We identify $p$ with the diagonal matrix
$p\equiv \diag(p_1,\ldots, p_n)$
and choose the normalization of the natural symplectic form
$\Omega_{T^* \bT_n^0}$ as
\be
\Omega_{T^* \bT_n^0}\equiv \sum_{k=1}^n dp_k \wedge d q_k =
\frac{1}{2} \Im \tr (dp \wedge e^{-2\ri q} d e^{2\ri q}).
\label{3.18}\ee
The point is that $\bT_n^0$ is a covering space of $Q(n)$, and $T^* \bT_n^0$ is a symplectic covering space
of $T^* Q(n)$.   To state this formally, we have
\be
\pi_1 ^* (\Omega_{T^* Q(n)}) = \Omega_{T^* \bT_n^0},
\label{3.19}\ee
where
\be
\pi_1:T^* \bT_n^0 \to  (T^*\bT_n^0)/ S(n) \equiv T^* (\bT_n^0/S(n))\equiv T^* Q(n)
\label{3.20}\ee
is the natural submersion.
Therefore one can identify the Poisson algebra of the smooth
functions on $T^* Q(n)$ with the Poisson algebra of the smooth $S(n)$-invariant functions
on $T^* \bT_n^0$. For example, one may regard the Lax matrix $L(q,p)$ (\ref{1.1})
as a function on $T^* \bT_n^0$,
with its spectral invariants (symmetric functions) giving the commuting Hamiltonians on $T^* Q(n)$.
Actually  $Q(n)$ is a rather non-trivial manifold,
and the ensuing technical complications are avoided if one works with the $S(n)$-invariant functions
on $T^* \bT_n^0$.

\medskip
 \noindent {\bf Theorem 3.1.}
{\it Consider the smooth map $\tilde \I: T^* \bT_n^0 \to GL(n,\bC)$ defined by the formula
\be
\tilde \I(e^{2\ri q},p)_{kk}=e^{ -\frac{p_k}{2}-2\ri q_k}  \prod_{m<k}\left[1 +
\frac{\sinh^2\frac{x}{2}} { \sin^2(q_k - q_m)}  \right]^{-\js}\prod_{m>k} \left[1 +
\frac{\sinh^2\frac{x}{2}} { \sin^2(q_k - q_m)}  \right]^{\js},
\label{Seq1}\ee
\be
\tilde  \I(e^{2\ri q},p)_{kl}=0, \quad k>l, \qquad
\tilde\I(e^{2\ri q},p)_{kl}= \tilde \I(e^{2\ri q},p)_{ll}\prod_{m=1}^{l-k}
\frac{e^{\frac{x}{2}}e^{2\ri q_l}-e^{-\frac{x}{2}}
 e^{2\ri q_{k+m}}}{e^{2\ri q_l}- e^{2\ri q_{k+m-1}}},
\quad  k<l,
\label{Seq2}\ee
where $T^* \bT_n^0$ is parametrized according to (\ref{3.17}).
This map is  injective, its image lies in $F_{\nu(x)}$, and it enjoys the property
\be
\tilde \I^* (\omega_+) = \Omega_{T^* \bT_n^0}.
\ee
Moreover, $\tilde \I$ descends to  a diffeomorphism
$\I: T^*Q(n) \to F_{\nu(x)}/{\bar G}_{\nu(x)}$ defined by the equality
\be
\I \circ \pi_1 = \pi \circ \tilde \I,
\label{3.24}\ee
with the  projections $\pi_1: T^* \bT_n^0 \to T^*Q(n)$,
$\pi: F_{\nu(x)} \to F_{\nu(x)}/{\bar G}_{\nu(x)}$, and
there holds the relation
\be
\I^*(\Omega_{\nu(x)}) = \Omega_{T^* Q(n)}.
\ee
Thus $(P,\omega)$  (\ref{3.16}) is a model
of the reduced phase space $(F_{\nu(x)}/\barG_{\nu(x)},\Omega_{\nu(x)})$.
}

\medskip
\noindent
{\bf Remark 3.3.} The image of $T^* \bT_n^0$ by $\tilde \I$
 is a symplectic submanifold $\tilde S  \subset GL(n,\bC)$.
$(\tilde S, \omega_+\vert_{\tilde S})$ and  $T^*\bT_n^0$ are  symplectomorphic by $\tilde \I$, and
furnish symplectic covering spaces of the reduced phase space.

\medskip
  \noindent {\bf Theorem 3.2.}
{\it   The  composition map ${\cal L}\circ \tilde \I$,
with $\cL$ in (\ref{Unlax}),
gives (up to an inessential similarity transformation)
the Lax matrix  $L$ (\ref{1.1}) of the original Ruijsenaars-Schneider system, where $L$ is regarded
as a function on
the covering space $T^* \bT_n^0$ of $P= T^* Q(n)$.}

\medskip

The previous Theorems 3.1 and 3.2 were essentially obtained in  our letter \cite{LMP}, but
there the covering space aspects and some other non-trivial details were omitted.
We shall give full proofs of them later in this paper.
 The  forthcoming Theorems 3.3  and 3.4  encapsulate the main original results obtained in this paper.
 They are  technically  more involved than Theorems 3.1 and 3.2  and, in  order to formulate them,
 we first need to define several    matrix-valued functions of various sets of arguments. The reader
 who does not wish to be slowed down in following the
 conceptual line of presentation  may   skip   the details of these  definitions
 on first reading  and continue directly to the statements of the theorems afterwards.

\noindent  {\bf Definition 3.1.}
\emph{For any $x\in\bR$, define the following subsets of $\bR^n$:}
  \be \C_x:=\{\hat p\in \bR^n \,\vert\,
\hat p_l-\hat p_{l+1}> \frac{\vert x\vert}{2}, \, 1\leq l<n\},
  \,\,\,
  \bar\C_x:=\{\hat p\in\bR^n\,\vert\, \hat p_l-\hat p_{l+1}\geq
   \frac{\vert x\vert}{2},  \,  1\leq l<n\}.
   \label{gap} \ee

   \medskip

  \noindent {\bf Remark 3.4.}  We shall refer  to the elements of $\C_x$ as   the interior  elements
  of $\bar\C_x$.
  Note that $\bar\C_0$ is the standard Weyl chamber associated with $gl(n,\bC)$ and
the phase space of the system
$(\hat P, \hat \omega, \hat L)$  is $\hat P= \bT_n \times \C_x$.
We often identify an element of $\bR^n$ with a corresponding $n\times n$  diagonal matrix.
For example, $\hat q$ in (\ref{1.4}) parametrizes $\bT_n$ by $e^{\ri \hat q}$ and we may use
$\hat p\simeq \diag(\hat p_1,\ldots, \hat p_n)$.

\medskip

\noindent{\bf Definition 3.2.} {\it   Consider the phase space $\hat P=\bT_n \times \C_x$ (\ref{1.4})
  of the dual Ruijsenaars-Schneider system
   and the symplectic manifold $\hp:=  \bC^{n-1}\times \bC^\times$, where  $\bC^\times$  denotes
 the complex plane without the origin and   the symplectic form $\hat\om_c$
  on $\hp$  is defined by
\be \ho:= \frac{\ri dZ\w d\bar Z}{2\bar ZZ} +{\mathrm{sign}}(x)\sum_{j=1}^{n-1}\ri dz_j\w d\bar z_j,
\qquad Z\in \bC^\times, \quad z\in \bC^{n-1}.
\label{hatom}\ee
  Define the  smooth injective map  $\Z_x:\hat P\to \hat P_c$  by
 \be
 z_j(x,\hat q,\hat p)=(\hat p_j-\hat p_{j+1}-x/2)^{\jp}\prod_{k=j+1}^ne^{-\ri\hat q_k},
 \,\, j=1,..., n-1, \,\, Z(x,\hat q,\hat p)=e^{-\hat p_1} \prod_{k=1}^ne^{-\ri\hat q_k},\,\,  x>0,
  \label{x+}\ee
 \be
 z_j(x,\hat q,\hat p)=(\hat p_j-\hat p_{j+1}+x/2)^{\jp}\prod_{k=1}^je^{-\ri\hat q_k},
 \,\, j=1,..., n-1, \,\,
 Z(x,\hat q,\hat p)=e^{-\hat p_n}\prod_{k=1}^ne^{-\ri\hat q_k},\,\, x<0.\label{x-}\ee}

 \medskip
\noindent {\bf Remark 3.5.} One can check that $\Z_x$ is a
symplectic embedding of $(\hat P, \hat \omega)$ into $(\hat P_c,
\hat \omega_c)$, i.e., \be \Z_x^* \hat \omega_c = \hat \omega
\label{SRembed}\ee with $\hat \omega$ in (\ref{1.4}). The
$\Z_x$-image of $\hat P$  in $\hat P_c$ is a dense open submanifold;
$\hat P_c\setminus \Z_x(\hat P)$ consists of the points for which one
or more of the complex coordinates $z_j$ is equal to zero. It is
important to note that the same embedding of $\hat P$ into $\hat
P_c$  was also used in \cite{RIMS95}\footnote{One can see this from
Eq.~(1.73) in \cite{RIMS95}, where the completion of the dual phase
space was formulated in terms of a covering space of $\hat P$.}.
This fact, together with the requirements forced on us
by the technical analysis in Section 5,
 motivated  Definition 3.2.

\medskip
 \noindent{\bf Definition 3.3.} {\it
 By introducing the `special index'
  $a:=n$ for $x>0$ and  $a:=1$ for $x<0$,
  define the $n\times n$ orthogonal matrix  $\kappa_L(x)$ by
\be
\kappa_L(x)_{aa}=\frac{v_a(x)}{\sqrt{n}},\,\,
\kappa_L(x)_{ij}=\delta_{ij}-\frac{v_i(x)v_j(x)}{n+\sqrt{n}v_a(x)},\,\,
\kappa_L(x)_{ia}= -\kappa_L(x)_{ai}=\frac{v_i(x)}{\sqrt{n}},\,\,
i,j\neq a,
\label{C}\ee
where  $v(x)$ is given by (\ref{vx}).
Then consider $z\in \bC^{n-1}$ and introduce the smooth functions
\be Q_{jk}(x,z)=
 \sqrt{\frac{\sinh{( \sum_{l=j}^{k-1}z_l\bar z_l +(k-j)\fv -\fp)}}
 {\sinh{(  \sum_{l=j}^{k-1}z_l\bar z_l +(k-j)\fv )}}}, \quad 1\leq j<k\leq n,
 \ee
 with  $Q_{jk}(x,z):=Q_{kj}(-x,z)$ for $j>k$.
By using  the above notations and $J(y):= \sqrt{\frac{\sinh{y} }{y}}$ for $y\neq 0$,
$J (0):= 1$,
define  the  smooth  $n\times n$ matrix function $\hat\zeta(x,z)$ as
\be
\hat\zeta(x,z)_{aa}=\sqrt{\frac{\sinh{\frac{x}{2}}}{\sinh{\frac{nx}{2}}}}\prod_{l\neq a}Q_{al}(x,z),
\quad
\hat\zeta(x,z)_{aj}=-\overline{\hat\zeta(x,z)_{ja}},
\quad j\neq a,
\ee
\be
 \hat\zeta(x,z)_{jn}=
 \sqrt{\frac{\sinh{\frac{x}{2}}}{\sinh{\frac{nx}{2}}}}\frac{z_jJ(z_j\bar z_j)}
 {\sqrt{\sinh{(z_j\bar z_j+\fp)}}}\,\,\prod_{l\neq j,j+1}Q_{jl}(x,z), \qquad x>0, \quad j\neq n,
 \label{aca}\ee
\be
\hat\zeta(x,z)_{j1}= \sqrt{\frac{\sinh{\frac{x}{2}}}{\sinh{\frac{nx}{2}}}}\frac{z_{j-1}
J(z_{j-1}\bar z_{j-1})}{\sqrt{\sinh{(z_{j-1}\bar z_{j-1}-\fp)}}}\,\,\prod_{l\neq j-1,j}Q_{jl}(x,z),
\qquad x<0, \quad j\neq 1,
\ee
\be
\hat\zeta(x,z)_{jk}=\delta_{jk}+\frac{\hat\zeta(x,z)_{ja}\hat\zeta(x,z)_{ak}}{1+\hat\zeta(x,z)_{aa}},
\qquad j,k\neq a.
 \label{sfc}\ee
Next, define the smooth $n\times n$  matrix function $\hat\theta(x,z)$   for $x>0$ as
\be
\hat\theta(x,z)_{jk}=
\frac{(\sinh{\frac{nx}{2}}){\rm sign}(k-j-1)\hat\zeta(x,z)_{jn}\hat\zeta(-x,z)_{1k}}
{\sinh{(\sum_{l={\rm min}(k,j)}^{{\rm max}(k,j)-1}z_l\bar z_l +\vert k-j-1\vert \frac{ x }{2} )}},
\quad   k\neq j+1, \label{hth1}\ee
 \be
 \hat\theta(x,z)_{j,j+1}=\frac{-\sinh{\fp}}{\sinh{(z_j\bar z_j+\fp)}} \prod_{l\neq j,j+1}
 Q_{jl}(x,z)Q_{j+1,l}(-x,z),
 \label{hth2}\ee
and for $x<0$ as
\be
\hat\theta (x,z)=\hat\theta(-x,z)^\dagger.
\label{hth3}\ee
Finally, let $\Delta(x,z,Z)$ be the diagonal matrix  function on $\hat P_c$ given
for $x>0$
by the components
\be
\Delta_1(x,z, Z)=Z, \qquad
 \Delta_j(x,z,Z)= \vert Z\vert\exp{(\sum_{l=1}^{j-1}z_l\bar z_l +(j-1)\fp )},    \quad j=2,...,n,
 \label{D}
 \ee
 and for $x<0$  by the components
\be
\Delta_n(x,z,Z)=Z, \qquad
\Delta_j(x,z,Z)= \vert Z\vert\exp{(-\sum_{l=j}^{n-1}z_l\bar z_l +(n-j)\fp )},
 \quad j=1,...,n-1.
  \label{D'}\ee
 }

\medskip
\noindent
{\bf Remark 3.6.}
 The origin of the above formulae  will become clear in Section 5, and there
we shall demonstrate the unitarity
  of the matrices  $\hat\theta(x,z)$ and $\hat\zeta(x,z)$  for all
values of their arguments.

Recall Remark 3.2 concerning our terminology for a global cross section.

\medskip
\noindent {\bf Theorem 3.3.} {\it
The symplectic manifold $(\hp,\ho)$ (\ref{hatom}) is a model of the reduced phase space
$(F_{\nu(x)}/\barG_{\nu(x)},\Omega_{\nu(x)})$.
With the notations introduced in Definition 3.3,
the map $\hat \I: \hat P_c \to F_{\nu(x)}$ given by
\be
\hat \I(z,Z):= \biggl(\kappa_L(x)\hat\zeta(x,z)^{-1}\biggr)\trl\biggl(\Delta(x,z,Z)\hat\theta(x,z)^{-1}\biggr)
\label{dif}\ee
is a global cross section,
and $\pi \circ \hat \I: \hat P_c \to F_{\nu(x)}/\bar G_{\nu(x)}$ is a symplectomorphism.
}

\medskip
   \noindent {\bf Theorem 3.4.} {\it By using the symplectic
   embedding $\Z_x:\hat P\to \hp$ introduced in Definition 3.2 and $\hat \cL$ defined in (\ref{Unlax}),
   the  composition map $\hat {\cal L}\circ \hat \I  \circ\Z_x$
   gives (up to an inessential similarity
   transformation) the Lax matrix  $\hat L$ (\ref{1.3}) of the dual
   Ruijsenaars-Schneider system $(\hat P, \hat \omega, \hat L)$. }

 \medskip

To sum up,  Theorems 3.1 and 3.2 state that the original trigonometric
Ruijsenaars-Schneider system $(P,\omega, L)$ is exactly the result of the
 symplectic reduction of the  canonical free system $(\gln,\omega_+,{\cal L})$.
 Theorems 3.3 and 3.4 affirm that the  reduction of the other canonical free system
  $(\gln,\omega_+,\hat {\cal L})$  gives a certain  integrable system
  $(\hat P_c,\ho,\hat {\cal L}\circ \hat \I)$ which can be viewed
  (due to the non-surjectivity of the map $\Z_x:\hat P\to\hat P_c$)  as an
  extension  of the dual Ruijsenaars-Schneider system $(\hat P,\hat\om, \hat L)$.
  Moreover,
  together with Remark 3.5, they ensure
  that
  the extended  dual Ruijsenaars-Schneider system  $(\hat P_c,\ho,\hat {\cal L}\circ \hat \I)$
  coincides with the `minimal completion' of the system $(\hat P, \hat \omega, \hat L)$
  constructed  by Ruijsenaars by means of the direct
   method \cite{RIMS95}.
 The guiding principle behind his extension of $\hat P$
was the aim to obtain a \emph{bijective}
  correspondence between the phase spaces of the dual pair of systems.
 At the same time, the extension
 gave rise to the completion of the dual flows, which are
 not complete on $\hat P$.
  It is pleasing that  Ruijsenaars'  minimal completion  comes about naturally  from
   the symplectic reduction. In this framework the geometric origin of the
   duality symplectomorphism
   between $(P, \omega)$ and $(\hat P_c, \hat \omega_c)$,
which has been established in \cite{RIMS95} by a complicated web of arguments,
    becomes transparent:
    any
   two models of the   reduced phase space are naturally
   symplectomorphic to each other. The natural
   symplectomorphism maps to each other those points of the two different models
   that correspond to the same point of the reduced phase space.

We shall further discuss in Section 6  why the geometrically induced
symplectomorphism between $(P, \omega)$ and $(\hat P_c,\ho)$ is the same as
the duality map (alias `action-angle map') constructed in \cite{RIMS95}.

\section{  Proofs of Theorems 3.1 and 3.2}
\setcounter{equation}{0}

Many ingredients  of the proofs that follow were already given in our previous paper
\cite{LMP} but here we present all this  material in a more complete and natural way. In
particular,
we shall explain how the non-trivial topology of the
configuration space  of indistinguishable particles on the circle (see also Appendix B) is reflected in
the symplectic reduction and  why this aspect of the story explains
the geometric origin of the important formula (3.13) of \cite{LMP}. We recall that  the
somewhat complicated  formula
(3.13) of \cite{LMP} (which appears as (\ref{N.20}) below)
relates the group theoretically simplest coordinates on the
reduced phase space  with the cotangent-bundle coordinates in which the
Ruijsenaars-Schneider Hamiltonians are usually expressed.

Let $A<B$ denote the subgroup  of diagonal matrices with positive real entries and
 $N < B$ the subgroup of upper-triangular
matrices with unit diagonal.
Define the smooth function $\cN: \bT_n^0 \to N$ by the formula
\be
 \cN(T)_{kl}=\prod_{m=1}^{l-k}
\frac{e^{\frac{x}{2}}T_l -e^{-\frac{x}{2}}
 T_{k+m}}{T_l - T_{k+m-1}},
\qquad \forall k<l,
\label{N.1}\ee
and introduce the subset $\tilde S\subset GL(n,\bC)$ as follows:
\be
\tilde S:= \{ \cN(T) a T^{-1}\,\vert\, a\in A,\, T\in \bT_n^0\}.
\label{N.2}\ee

\medskip
\noindent
{\bf Lemma 4.1.} \emph{The set
$\tilde S$ lies in the constraint-manifold $F_{\nu(x)} \subset GL(n,\bC)$ and
it  intersects every orbit of the gauge group
$\bar G_{\nu(x)}$ acting on $F_{\nu(x)}$.
Every $K\in F_{\nu(x)}$ with $\Xi_R(K)\in \bT_n$ belongs to $\tilde S$.
The map
\be
\bT_n^0 \times A \to F_{\nu(x)},
\quad
(T,a) \mapsto  \cN(T) a T^{-1}
\label{N.3}\ee
is an embedding, and the corresponding pull-back of the form $\omega_+$ is the symplectic form
\be
\omega_{\tilde S}= \Im\tr(T^{-1} dT \wedge a^{-1} da).
\label{N.4}\ee
}

\medskip
 \noindent
 {\bf Proof.}
 The statement is just a reformulation of part of Theorem 1 of \cite{LMP}.
 The fact that $\tilde S$ is an embedded submanifold of $GL(n,\bC)$ is obvious from
 the Iwasawa decomposition, which also implies by Lemma 3.2 that $\tilde S$ is an embedded submanifold
 of $F_{\nu(x)}$.
\emph{Q.E.D.}

\medskip
\noindent
{\bf Lemma 4.2.} \emph{For every fixed $K\in \tilde S$ and permutation $\sigma \in S(n)$
there exists a unique element $[g(K,\sigma)]\in \bar G_{\nu(x)}$ for which
\be
\Psi( [g(K,\sigma)], K) \in \tilde S
\quad\hbox{and}\quad
\Xi_R(\Psi( [g(K,\sigma)], K)) = \sigma(\Xi_R(K)),
\label{N.5}\ee
where $\Psi$ denotes the action (\ref{identify2})
and $\sigma(T)$ is obtained by permuting the entries of any $T\in \bT_n^0$.
All gauge transformations that map $K\in \tilde S$ to $\tilde S$ are of the above type,
and the formula
\be
\Phi: S(n) \times \tilde S \to \tilde S,
\quad
\Phi_\sigma(K):=\Phi(\sigma,K):= \Psi([g(K, \sigma)], K)
\label{N.6}\ee
defines a smooth, free action of $S(n)$ on $\tilde S$, which preserves the symplectic form $\omega_{\tilde S}$.
}

\medskip
 \noindent
 {\bf Proof.}
 This can be extracted from \cite{LMP}, too, and thus we can be brief here.
 First, for fixed $K\in \tilde S$ and $\sigma\in S(n)$ there cannot
 exist two gauge transformations subject to (\ref{N.5}),
 since the action of $\bar G_{\nu(x)}$ is free on $F_{\nu(x)}$ and two different elements $K_1, K_2 \in \tilde S$
 satisfying $\Xi_R(K_1)=\Xi_R(K_2)$ are never gauge equivalent \cite{LMP}.
 Second, because of the second relation in (\ref{2.33})
 and the surjectivity of the map $g_K:U(n)\to U(n)$ given by
 $g_K(\eta)=\Xi_R(\eta\Lambda_L(K))$ for each $K$, it is clear that for any $K\in \tilde S$ and
  $\sigma\in S(n)$
 there exists some $\eta\in G$ for which
\be
\Xi_R(\eta \triangleright  K) = \sigma(\Xi_R(K)).
\label{N.7}\ee
This implies that
\be
(\Dress_\eta (\nu(x)))_{jj} = \Lambda(\eta \triangleright K)_{jj}=1,
\qquad
\forall j,
\label{N.8}\ee
where we used both the equivariance of the moment map as well
the formula (\ref{2.22}), which shows that
$\Lambda(K)_{jj}=1$ if $K = b T^{-1}$ for some $b\in B$ and $T \in \bT_n$.
Next, it is not difficult to see  (e.g.~from the proof of Lemma 1 in \cite{LMP}) that for any element
$\Dress_\eta (\nu(x))$ with unit diagonal there exists some $\tau \in \bT_n$
for which  $\Dress_\eta (\nu(x))= \Dress_\tau(\nu(x))$.
Then it follows from (\ref{N.7}) that $\Xi_R(\tau^{-1}\eta \triangleright  K) = \sigma(\Xi_R(K))\in \bT_n$.
Consequently,
$[g(K, \sigma)]:= [ \tau^{-1} \eta]\in \bar G_{\nu(x)}$ is the required element.

Each gauge transformation that maps $K\in \tilde S$ to $\tilde S$ is associated with some $\sigma \in S(n)$
according to (\ref{N.5}),
since these gauge transformations act by some permutation on $\Xi_R(K)$ (again because of (\ref{2.33})).
It is clear from the established uniqueness property that (\ref{N.6}) defines indeed a smooth,
free action of $S(n)$ on $\tilde S$. This action preserves $\omega_{\tilde S}$ since it is given by gauge
transformations
(the gauge transformations preserve $\omega_{+}\vert_{F_\nu(x)}$, and $\omega_{\tilde S}$ is the pull-back
of $\omega_{+}\vert_{F_\nu(x)}$ on $\tilde S$).
 \emph{Q.E.D.}
\medskip

The following important formula was found by first making a detailed inspection in the $n=2$ case,
and then generalizing the result for arbitrary $n$.

 \medskip
\noindent
{\bf Lemma 4.3.} \emph{The action of the transposition $\sigma_{k,k+1}\in S(n)$
on $\cN(T) a T^{-1} \in \tilde S$ is given explicitly by the formula
\be
\Phi_{\sigma_{k,k+1}}(\cN(T) a T^{-1}) = \cN(\sigma_{k,k+1}(T))\hat a \sigma_{k,k+1}(T)^{-1},
\label{N.9}\ee
where $\hat a_j = a_j$ if $j\notin\{ k, k+1\}$ and
\be
\hat a_k = a_{k+1} W_k(T),
\,\,\,
\hat a_{k+1} = \frac{a_k }{W_k(T)}
\,\,\,\,
\hbox{with}\,\,\,\,
W_k(T):=
\left[ 1 + \frac{\sinh^2\frac{x}{2}}{\sin^2(q_k-q_{k+1})}\right]^{\frac{1}{2}},
\,\,\,  T=e^{2 \ri q}.
\label{N.10}\ee
}

\medskip
 \noindent
 {\bf Proof.}
Fix $1\leq k\leq (n-1)$ and $K= \cN(T) a T^{-1} \in \tilde S$.
For any $\gamma \in (0,\pi)$ define the matrix  $g(\gamma)\in SU(2)$  by
\be
g(\gamma):= \left[\begin{matrix} \alpha & \beta\\ \beta & \bar\alpha \end{matrix}\right],
\qquad
\alpha:= (\cos\gamma + \ri \sin \gamma \tanh \frac{x}{2}),\quad
\beta:= {\ri \sin\gamma}/{\cosh\frac{x}{2}}.
\label{N.11}\ee
Then introduce the element $g_k(\gamma) \in U(n)$ by
\be
g_k(\gamma):= \diag(\Gamma,\ldots, \Gamma, g(\gamma), \Gamma,\ldots, \Gamma),
\quad
\Gamma:= e^{\ri \gamma},
\label{N.13}\ee
where the first string of $\Gamma$'s occupies the first $(k-1)$-entries along the diagonal.
Introduce similarly   the matrix $\chi_k(\gamma) \in U(n)$ by
\be
\chi_k(\gamma):= \diag(\Gamma, \ldots, \Gamma, \chi, \Gamma, \dots, \Gamma)
\quad
\hbox{with}\quad
\chi:=\left[\begin{matrix} 0 & \ri \\ \ri & 0 \end{matrix}\right].
\label{N.14}\ee
It is readily verified that the vector $v(x)$ (\ref{vx}) is an eigenvector of $g_k(\gamma)$,
\be
g_k(\gamma) v(x) = e^{\ri \gamma} v(x),
\label{N.15}\ee
which implies that $g_k(\gamma) \triangleright \nu(x) = \nu(x)$, i.e., $[g_k(\gamma)]$ belongs to the gauge group
$\bar G_{\nu(x)}$.
It is also straightforward to check that if $\gamma$ is determined by the equality
\be
\cot \gamma = (\tanh x/2) \cot(q_k - q_{k+1}),
\qquad
T_j = e^{2\ri q_j}
\quad (\forall j=1,\ldots, n),
\label{N.16}\ee
then the following Iwasawa decomposition is valid:
\be
g_k(\gamma) \cN(T) a = b \chi_k(\gamma)
\quad\hbox{with}\quad b \in B,\quad
b_{j,j}= \hat a_j
\quad (\forall j=1,\ldots, n),
\label{N.17}\ee
where $\hat a \in A$ is as claimed by the lemma.
To finish the proof, we notice from (\ref{N.17}) that
$\Xi_R( g_k(\gamma) \Lambda_L(K)) = \chi_k(\gamma)^{-1}$, and this allows us to calculate (cf.~(\ref{2.33}))
that
\be
\Xi_R ( g_k(\gamma) \triangleright K) = \sigma_{k,k+1}(T).
\label{N.18}\ee
By the second sentence of Lemma 4.1, this implies that $g_k(\gamma) \triangleright K\in \tilde S$.
Therefore the element $[g(K, \sigma_{k,k+1})]$ of Lemma 4.2 is provided by $g_k(\gamma)$ with
$\gamma$ in (\ref{N.16}).
Since any $K \in \tilde S$ is determined by $\Xi_R(K)$ and the diagonal part of $\Lambda_L(K)$,
it follows that $b$ in (\ref{N.17}) is given by $b= \cN(\sigma_{k,k+1}(T)) \hat a$,
which can be checked also by direct calculation.
 \emph{Q.E.D.}

\medskip
\noindent
{\bf Lemma 4.4.}
\emph{The image of the map $\tilde \I$ defined in Theorem 3.1 is
the submanifold $\tilde S$ defined in (\ref{N.2}).
The corresponding map $\tilde \I: T^* \bT_n^0 \to \tilde S$ is an $S(n)$-equivariant
symplectic diffeomorphism.
Here, we refer to the $S(n)$-action on $(T^* \bT_n^0, \Omega_{T^* \bT_n^0})$ obtained
as the cotangent lift of the permutation action on $\bT_n^0$ and to the $S(n)$-action
on $(\tilde S, \omega_{\tilde S})$ described in Lemmas 4.2 and 4.3.
}

\medskip
 \noindent
 {\bf Proof.}
It follows from the definition of $\tilde \I$ in Theorem 3.1 and from equation (\ref{N.1})
that we have the equality
\be
\tilde\I(e^{2\ri q},p) = \cN(e^{2\ri q}) a(e^{2\ri q},p) e^{-2\ri q}
\label{N.19}\ee
with the function $a= \diag(a_1,\ldots, a_n)$ given by
\be
a_j(e^{2\ri q},p): = e^{-\frac{p_j}{2}}
 \prod_{m<j} \left[ 1 + \frac{\sinh^2\frac{x}{2}}{\sin^2(q_j - q_m)}\right]^{-\frac{1}{4}}
 \prod_{m>j} \left[ 1 + \frac{\sinh^2\frac{x}{2}}{\sin^2(q_j - q_m)}\right]^{\frac{1}{4}},
 \quad
j=1,\ldots, n.
\label{N.20}\ee
By taking into account the identification (\ref{3.17}) and the definition (\ref{N.2}), this ensures the validity
of the first sentence of the lemma. One can see from the formula of $\tilde \I$ or directly from Lemma 4.1
that $\tilde \I: T^* \bT_n^0 \to \tilde S$ is a diffeomorphism.
The symplectic property $\tilde \I^* \omega_{\tilde S} = \Omega_{T^* \bT_n^0}$ (with (\ref{3.18}) and (\ref{N.4}))
can be established directly. In fact, the properties mentioned so far would hold also if one replaced
$[1 + (\sinh^2\frac{x}{2})/\sin^2(q_j - q_m)]$ in (\ref{N.20}) by any positive even function
$W(q_j - q_m)$.

It is sufficient to confirm the equivariance property of $\tilde \I$ for the transpositions
\be
\sigma:=\sigma_{k,k+1}\in S(n),
\quad k=1,\ldots, n-1.
\label{N.21}\ee
From (\ref{N.19}) we obtain
\be
\tilde\I(\sigma(e^{2\ri q}),\sigma(p)) = \cN(\sigma(e^{2\ri q}))
a(\sigma(e^{2\ri q}),\sigma(p)) (\sigma(e^{2\ri q}))^{-1}.
\label{N.22}\ee
For any fixed $k$, is easily checked that
$a_j(\sigma(e^{2\ri q}),\sigma(p))= a_j(e^{2 \ri q}, p)$ if $j\notin\{k,k+1\}$ and
\be
a_k(\sigma(e^{2\ri q}),\sigma(p)) = a_{k+1}(e^{2\ri q},p) W_k(e^{2\ri q}),
\quad
a_{k+1}(\sigma(e^{2\ri q}),\sigma(p)) = \frac{a_k(e^{2\ri q},p)}{ W_k(e^{2\ri q})}
\label{N.23}\ee
with the same function $W_k$ as in (\ref{N.10}).
The comparison of (\ref{N.22}) with (\ref{N.9}) shows that the proof is complete.
 \emph{Q.E.D.}

Our lemmas explain the geometric picture
behind Theorem 3.1, which is now easy to prove.

 \medskip
\noindent
{\bf Proof of Theorem 3.1:}
 Lemmas 4.1 and 4.2 give rise to the identification of  symplectic manifolds
\be
\tilde S/S(n) \simeq F_{\nu(x)}/{\bar G_{\nu(x)}}.
\label{N.24}\ee
Here, $\tilde S/S(n)$ is the space of orbits of the $S(n)$-action given by Lemma 4.2,
its symplectic form descends from $\omega_{\tilde S}$ on $\tilde S$,
while $(F_{\nu(x)}/{\bar G_{\nu(x)}}, \Omega_{\nu(x)})$ is the reduced phase space of interest.
Moreover, we constructed the following commutative diagram of maps:
\be\begin{array}{ccc}
T^* \bT_n^0 &
\ \stackrel{{\tilde{\I} }}{\longrightarrow}\ & \tilde S \\
 {} & {} &\\
\pi_1 \,\, \downarrow {\phantom{\scriptstyle{{\cal D}_0 \times \rm{id}}}}& {} &  {\phantom{\scriptstyle{{\cal D}}}} \downarrow \,\, {\pi_{\tilde S}} \\
{} & {} & \\
 T^*Q(n) & \ \stackrel{{\I}} \longrightarrow\  & \tilde S/S(n)\\
\end{array}\label{diagram}\ee
The map $\I$ is well-defined by this diagram and is a  diffeomorphism, because of Lemma 4.4.
(To compare with (\ref{3.24}), note that $\pi_{\tilde S} \circ \tilde \I = \pi\circ \tilde \I$
since the image of $\tilde \I$ is $\tilde S \subset F_\nu(x)$.)
We also established the relation (\ref{3.19}) as well as $\pi_{\tilde S}^* (\Omega_{\nu(x)}) = \omega_{\tilde S}$
(by (\ref{N.24})) and
$\tilde \I^*( \omega_{\tilde S})= \Omega_{T^* \bT_n^0}$ (by Lemma 4.4).
These relations and the fact that $\pi_1$ and $ \pi_{\tilde S}$ are
local diffeomorphisms imply that $\I^* (\Omega_{\nu(x)}) = \Omega_{T^* Q(n)}$.
\emph{Q.E.D.}

 \medskip
\noindent
{\bf Proof of Theorem 3.2:}
Denote by $\cL^{\tilde S}$ the restriction of the unreduced Lax matrix $\cL$ (\ref{Unlax}) to $\tilde S$
(\ref{N.2}).
The definition directly yields the formula
\be
\cL^{\tilde S} = T a^{-1} \cN(T)^{-1}  (\cN(T)^{-1})^\dagger a^{-1} T^{-1},
\label{C2}\ee
where $T$, $a$ and $\cN(T)$ are understood as evaluation functions on $\tilde S \simeq \bT_n^0 \times A$.
Next, we remark that the restriction of the moment map constraint (\ref{MMC}) to $\tilde S$ is equivalent
to the relation
\be
\cN(T) a \cL^{\tilde S} (\cN(T) a)^\dagger = \nu(x) \nu(x)^\dagger =
e^{-x}\left[ \1_n+ \frac{e^{nx}-1}{n}v(x)  v(x)^\dagger \right],
\label{C3}\ee
which can be rewritten as
\be
e^{\frac{x}{2}} \cL^{\tilde S}_{jk} - e^{-\frac{x}{2}} T_j^{-1} \cL_{jk}^{\tilde S} T_k =
 2 \cU_j \bar \cU_k a_j^{-1} a_k^{-1} \sinh\frac{x}{2}\,,
\label{C4}\ee
if we define
\be
\cU_j:= \left[\frac{e^{-\frac{x}{2}} (e^{nx} -1)}{2n \sinh\frac{x}{2}}\right]^{\frac{1}{2}}
\left( \cN(T)^{-1} v(x)\right)_j := \eta_j \vert \cU_j\vert.
\label{C5}\ee
Here, $T = \diag(T_1,\ldots, T_n)$ and $a= \diag(a_1,\ldots, a_n)$.
By solving  (\ref{C4}) for $\cL^{\tilde S}$ we arrive at
\be
\cL^{\tilde S}_{jk} = \eta_j
\frac{ 2 a_j^{-1}  \vert\cU_j\vert\,  a_k^{-1} \vert\cU_k\vert \sinh\frac{x}{2}}
{e^{\frac{x}{2}} - e^{-\frac{x}{2}} T_j^{-1} T_k} \eta_k^{-1}.
\label{C6}\ee
With the inverse  $\cN(T)^{-1}$ displayed in \cite{LMP}, it is also straightforward
to calculate that
\be
\vert \cU_j\vert =\prod_{m>j} \left[ 1 + \frac{\sinh^2\frac{x}{2}}{\sin^2(q_j - q_m)}\right]^{\frac{1}{2}},
\label{C7}\ee
where we use the parametrization $T_k=e^{2 \ri q_k}$ for all $k=1,\ldots, n$.
Next, let us parametrize $a\in A$ according to (\ref{N.20})
and insert also (\ref{C7}) into (\ref{C6}).
Then we obtain
\be
\cL^{\tilde S}_{jk} =\frac{\gamma_j \gamma_k^{-1}
e^{\frac{p_j+p_k}{2}}\sinh\frac{x}{2}}{\sinh(\frac{x}{2}+\ri q_j-\ri q_k)}
\prod_{m\neq j}\left[1 + \frac{\sinh^2\frac{x}{2}}{ \sin^2(q_j - q_m)}\right]^{\frac{1}{4}}
\prod_{s\neq k}\left[1 +
\frac{\sinh^2\frac{x}{2}} { \sin^2(q_k - q_s)}  \right]^{\frac{1}{4}}
\ee
with $\gamma_j:= \eta_j e^{\ri q_j}$.
Hence $\cL^{\tilde S}$  is conjugate to
the standard Ruijsenaars-Schneider
Lax matrix $L$ in (\ref{1.1}).

Recall from (\ref{N.19}) that the map
$\tilde \I: T^*\bT_n^0   \to F_{\nu(x)}$ in (\ref{Seq1}), (\ref{Seq2}) was obtained from the
map in (\ref{N.3}) by using the identification $T^*\bT_n^0\simeq  \bT_n^0 \times \bR^n$ (\ref{3.17})
and the parametrization (\ref{N.20}) of $a_j$.
Therefore the foregoing arguments prove Theorem 3.2.
\emph{Q.E.D.}

\medskip

\noindent
{\bf Remark 4.1.}
The statement of Theorem 3.2 was also obtained in \cite{LMP}, but there the details of the proof were omitted
for lack of space.
The `useful substitution'  (\ref{N.20}) was introduced in \cite{LMP} (Eq.~(3.13) in {\it loc.~cit.}) just on the basis that
it converts the expression $T a^{-1} \cN(T)^{-1}  (\cN(T)^{-1})^\dagger a^{-1} T^{-1}$ (\ref{C2})
into a conjugate of the standard Lax matrix $L(q,p)$ (\ref{1.1}).
The  deeper geometric meaning  of this substitution  is now revealed by Lemma 4.4  above.
Note that $\cL^{\tilde S}$ transforms by conjugation under the `residual gauge transformations'
of Lemma 4.2, and this corresponds to the fact that $L(q,p)$,
viewed as a function on $T^*\bT_n^0$ (\ref{3.17}),  transforms by conjugation
under the natural $S(n)$-action.

\medskip
\noindent
{\bf Remark 4.2.} The fact that $T^*Q(n)$ appears   as
a factor space  (\ref{diagram}), and not directly as  a global gauge slice,  reflects
the fact that $Q(n)$ is not a submanifold of $\bT_n^0$ (see also Appendix B).

\medskip
\noindent
{\bf Remark 4.3.}
It is worth pointing out that the consequence (\ref{C4}) of the moment map constraint is essentially
identical to the `commutation relation of the Lax matrix' that played an important r\^ole
in the analysis presented in \cite{RIMS95}.

\section{  Proofs of Theorems 3.3 and 3.4}
\setcounter{equation}{0}

   The proofs of Theorems  3.3 and 3.4   will be based on a series of preliminary
  lemmas. From now on we adopt the identification $\hat p\simeq \diag(\hat p_1,\ldots , \hat p_n)$ for any
  $\hat p\in \bar\C_0$ (\ref{gap}).

  \medskip

\noindent {\bf Lemma 5.1.} {\it  Every element $K$ of the group $GL(n,\bC)$
can be decomposed  as
\be K=k_L\trl (e^{-\hat p}k_R^{-1}),  \qquad k_L,k_R\in U(n), \quad \hat p\in \bar\C_0.
\label{Carm}\ee
Moreover, if  $\tau$ is any element  of  the maximal  torus $\bT_n < U(n)$ then  the triple
$k_L \tau, \hat p, \tau^{-1}k_R\tau$ gives the same element $K$ as the triple
$k_L,\hat p,k_R$,   and this is the maximal possible ambiguity of the decomposition (\ref{Carm})
if $\hat p$ is regular ($\hat p_i > \hat p_{i+1}$, $\forall i=1,...,n-1$).}

\medskip

\noindent {\bf Proof.}   The statement  of the lemma is a direct consequence
of the standard Cartan decomposition of the elements of $\gln$. Indeed,
it is well known that
 every element $K$ of the group $GL(n,\bC)$  can be   decomposed  as
\be
K=\eta_Le^{-\hat p}\eta_R^{-1}, \quad \eta_L,\eta_R\in U(n), \quad \hat p\in \bar\C_0.
\label{Car}\ee
For  each $K$, the diagonal  matrix $\hat p$  in the standard Cartan decomposition (\ref{Car})  is defined
unambiguously.  Moreover, simultaneous right multiplication of the pair  $\eta_L,\eta_R$  by an
element $\tau$
of the maximal
torus $\bT_n$ gives an equally good pair $ \eta_L\tau,\eta_R\tau$ and this is the maximal possible ambiguity
of the decomposition (\ref{Car}) if $\hat p$ is regular.

\medskip

  From (\ref{Car}) and the definition (\ref{2.21}) of the quasi-adjoint action, we obtain
\be
K=k_L\trl (e^{-\hat p} k_R^{-1}),  \qquad k_L,k_R\in U(n),
\ee
where
\be k_L=\eta_L,\qquad k_R=\Xi_R(\eta_L  e^{-\hat p})\eta_R.
\label{por}\ee
The proof is finished by noting that $(\eta_L \tau, \eta_R \tau)$
corresponds by  (\ref{por}) to $(k_L \tau, \tau^{-1} k_R \tau)$ for all $\tau\in \bT_n$.
\emph{Q.E.D.}

\medskip
\noindent
{\bf Lemma 5.2.} {\it  If $K=k_L\trl(e^{-\hat p}k_R^{-1})$ (\ref{Carm}) is a solution
of the moment map constraint (\ref{MMC}) then  $\hat p\in\bar\C_x$, as defined in (\ref{gap}).}

\medskip
\noindent
{\bf Proof.}
 By using the formula
\be
\Lambda_L(K)=\Lambda_R(K^{-1}), \quad \forall K\in \gln,
\ee
we can rewrite the moment map constraint (\ref{MMC}) as
\be
\Lambda(k_L\trl (e^{-\hat p} k_R^{-1}))=
\Lambda_L(k_Le^{-\hat p})\Lambda_L(\Xi_R^{-1}(k_Le^{-\hat p})k_Re^{\hat p})=\nu(x),
\ee
or, equivalently, as
\be
\Lambda(k_L\trl (e^{-\hat p} k_R^{-1}))\Lambda (k_L\trl (e^{-\hat p} k_R^{-1}))^\dagger=
k_Le^{-\hat p}k_Re^{2\hat p}k_R^{-1}e^{-\hat p}k_L^{-1}=\nu(x)\nu(x)^\dagger .
\ee
By means of (\ref{3.13}), the last equality can be further rewritten as
\be
k_Re^{2\hat p}k_R^{-1}=
\left[ e^{2\hat p}e^{-x}  + e^{-x}\frac{e^{nx}-1}{n}e^{\hat p}k_L^{-1}v(x)  v(x)^\dagger k_L e^{\hat p}\right].
\label{Jrov}\ee
The equality  of the characteristic polynomials of the matrices on the two sides of
(\ref{Jrov}) gives
\be
\prod_j(e^{2\hat p_j}-\lm)= \prod_j(e^{2\hat p_j-x}-\lm)  +
e^{-x}\frac{e^{nx}-1}{n}\sum_j \biggl(e^{2\hat p_j}\vert w_j\vert^2
\prod_{k\neq j}(e^{2\hat p_k-x}-\lm)\biggr),
\label{char}\ee
where
\be w:=k^{-1}_Lv(x)
\label{defw}\ee
 and $\lm$ is a complex variable.
To derive (\ref{char}),  we  used the identity
\be
\det(\1_n +uy^\dagger)=1+y^\dagger u,
\ee
which is valid for  arbitrary $n$-component column vectors  $u$ and $y$.

\medskip

Suppose that (\ref{Jrov}) holds  for some  regular $K$, i.e., $\hat p_1 > \hat p_2 >...> \hat p_n$.
We can then evaluate
 the polynomials on  both sides of (\ref{char}) at the $n$ different values
 $\lm=e^{2\hat p_j-x}, j=1,...,n$.
 This  yields
 \be
 \vert w_j\vert^2=n\frac{1-e^{-x}}{1-e^{-nx}}
 \prod_{k\neq j}\frac{1-e^{2\hat p_j-2\hat p_k-x}}{1-e^{2\hat p_j-2\hat p_k}},
 \qquad   j=1,\ldots,n.
 \label{eqw}\ee
 Consider first the case $x>0$.
 On account of $\hat p_1 > \hat p_2 >...> \hat p_n$,
  we find from (\ref{eqw})   for each $j=1,...,n-1$ the following inequality
 \be
 \prod_{k>j}(e^{2\hat p_j-2\hat p_k-x}-1)=
 \frac{1-e^{-nx}}{1-e^{-x}}\frac{\vert w_j\vert^2}{n} \prod_{k>j}(e^{2\hat p_j-2\hat p_k}-1)
 \prod_{k< j}\frac{1-e^{2\hat p_j-2\hat p_k}}{1-e^{2\hat p_j-2\hat p_k-x}}\geq 0.
 \label{ner}\ee
 Now we prove by induction that
  \be
  \hat p_{l}-\hat p_{l+1}-\frac{  x }{2}\geq 0,
  \qquad \forall l=1,...,n-1.
  \label{win}\ee
First of all, for $j=n-1$,  the inequality (\ref{ner})   gives immediately (\ref{win}) for $l=n-1$.
It is easy to see that if (\ref{win}) holds for   $l=j+1,j+2,...,n-1$,
 then  it holds also for $l=j$.
 Indeed, this follows from
  \be
  0\leq \prod_{k>j}(e^{2\hat p_j-2\hat p_k-x}-1)=
  (e^{2\hat p_j-2\hat p_{j+1}-x}-1)   \prod_{k>j+1}(e^{2\hat p_j-2\hat p_{k-1}+2\hat p_{k-1}-2\hat p_k-x}-1).
  \ee
  The case $x<0$ is very similar. The point of departure is the following inequality
  \be
  \prod_{k<j}(1-e^{2\hat p_j-2\hat p_k-x})=
  \frac{e^{-nx}-1}{e^{-x}-1}\frac{\vert w_j\vert^2}{n}
  \prod_{k<j}(1-e^{2\hat p_j-2\hat p_k})
  \prod_{k>j}\frac{e^{2\hat p_j-2\hat p_k}-1}{e^{2\hat p_j-2\hat p_k-x}-1}\geq 0,
  \quad \forall j=2,\ldots, n.\label{ner2}
  \ee
  In this case we prove by induction that
  \be
  \hat p_{l-1}-\hat p_{l}+\frac{  x }{2}\geq 0,
  \qquad \forall l=2,...,n.
  \label{win2}\ee
 Now  (\ref{ner2}) for $j=2$  gives (\ref{win2}) for $l=2$.
 By using
  \be
  0\leq \prod_{k<j}(1-e^{2\hat p_j-2\hat p_k-x})= (1-e^{2\hat p_j-2\hat p_{j-1}-x})
   \prod_{k<j-1}(1-e^{2\hat p_j-2\hat p_{k+1}+2\hat p_{k+1}-2\hat p_k-x}),
   \ee
 one sees that if (\ref{win2}) holds for   $l=2,...,j-1$, then  it holds also for $l=j$.

 So far we have proved Lemma  5.2 for the regular solutions of the moment map constraint
 (\ref{MMC}), i.e.,  for those $K$ (\ref{Carm})
  for which $\hat p_1 > \hat p_2 >...> \hat p_n$.
  We remark that such regular solutions exist. As an example, consider
  $K=k_L\trl (e^{-\hat p} k_R^{-1})$ with   $\hat p$ such that
 \be
 \hat p_l-\hat p_{l+1} =\frac{\vert x\vert}{2},
 \qquad \forall l=1,...,n-1.
 \label{A}
 \ee
 In fact, a solution is then provided by $k_L:=\kappa_L(x)$ (given by (\ref{C})) and  $k_R:=\kappa_R(x)$,
 where, for $x>0$
 \be
 \kappa_R(x)_{n1} =\kappa_R(x)_{i,i+1}(x)=1, \quad i=1,...,n-1,
 \qquad \kappa_R(x)_{ij}=0 \quad
 \hbox{otherwise,}
 \label{B}\ee
 and for $x<0$
\be \kappa_R(x):=\kappa_R(-x)^{-1}.
\label{B'}\ee

\medskip

 To finish the proof,  it remains to treat the case of  non-regular solutions of (\ref{MMC}),
  for which   two or more
  $\hat p_j$'s are equal to each other.
Suppose that such a non-regular solution, $K_0$,  exists.
Note that the space of solutions of (\ref{MMC}), $F_{\nu(x)}$, is connected, since it is the total
space of a principal fiber bundle with connected structure group and connected base,
as follows from Lemma 3.2 and Theorem 3.1.
Then take a regular solution, $K_1$, e.g.~the one exhibited above,  and connect
  $K_0$ with $K_1$ by a continuous path $K_s$, $s\in [0,1]$,
  in $F_{\nu(x)}$.
Now the
diagonal  matrix $\hat p(s)$  in  the  modified Cartan decomposition (\ref{Carm}) of $K_s$  (or, in other words,
the spectrum of the element $\sqrt{K_sK_s^\dagger}$) varies continuously with $s$. However, this is not possible
because the set of the non-regular elements of $\bar\C_0$ is
disconnected from $\bar\C_x$ for $x\neq 0$.
 Hence, non-regular solutions
  of the constraint (\ref{MMC}) do not exist.
  \emph{Q.E.D.}

\medskip
\noindent {\bf Lemma 5.3.}
{\it If $K=k_L\trl(e^{-\hat p}k_R^{-1})$ (\ref{Carm}) is a solution
of the moment map constraint (\ref{MMC}) then the matrix $k_R$ must have the form
\be
k_R=\tdl\theta(x,\hat p){\tdr}.
\label{tte}\ee
Here $\tdl,{\tdr}$ are some diagonal unitary matrices
and $\theta(x,\hat p)$
is the real orthogonal matrix defined for every  $\hat p\in \bar \C_x$
by
\be
\theta(x,\hat p)_{jk}:=
\frac{\sinh\left(\frac{x}{2}\right)}{\sinh\left(\hat p_k - \hat p_j  \right)}\prod_{m\neq j,k}
\left[\frac{\sinh(\hat p_j - \hat p_m - \frac{x}{2})\sinh(\hat p_k - \hat p_m +\frac{x}{2})}
{\sinh(\hat p_j - \hat p_m)
 \sinh(\hat p_k - \hat p_m)}\right]^{\frac{1}{2}},\quad j\neq k,
\label{5.23a}\ee
\be
\theta(x,\hat p)_{jj}:=\prod_{m\neq j}  \left[ \frac{\sinh(\hat p_j - \hat p_m - \frac{x}{2})\sinh(\hat p_j - \hat p_m +\frac{x}{2})}
{\sinh^2(\hat p_j - \hat p_m) }\right]^{\frac{1}{2}}.
\label{5.23b}\ee
}

\medskip
\noindent   {\bf Proof.}    Consider  the  following variant of the moment map constraint (\ref{Jrov}):
\be
k_R(x)e^{2\hat p} e^{\frac{x}{2}}k_R(x)^{-1}=e^{2\hat p}e^{-\frac{x}{2}}   +
2\left({\sinh}\frac{x}{2}\right) \xi(x)\xi(x)^\dagger,
\label{mod}\ee
where the vector $\xi(x)$ is defined as
\be
\xi(x):=\sqrt{\frac{e^{nx}-1}{n(e^x-1)} }e^{\hat p}k_L^{-1} v(x).
\label{jv}\ee
Here, our notation emphasizes the dependence of $k_R$ and $\xi$ on
 $x$ while their dependence on $\hat p$ remains tacit.
Observe from the comparison of (\ref{defw}), (\ref{eqw})  and (\ref{jv}) that
\be
\vert \xi(x)_j\vert^2=e^{2\hat p_j}
\prod_{k\neq j}\frac{e^x-e^{2\hat p_j-2\hat p_k}}{1-e^{2\hat p_j-2\hat p_k}}.
\label{norx}\ee
For any given $\hat p \in \bar\C_x$ and $\xi(x)$ subject to ({\ref{norx}), the constraint
(\ref{mod}) admits a
solution for
$k_R(x)$, since  the characteristic polynomials of the matrices on the two sides of
(\ref{mod}) are equal.
The solution $k_R(x)$ can be chosen to be unitary because
the matrix on the right hand side of ({\ref{mod}) is Hermitian.
Suppose that a pair $(k_R(x), \xi(x))\in U(n)\times \bC^n$ satisfies (\ref{mod}), at some fixed
$\hat p\in \bar\C_x$.
Then all pairs satisfying (\ref{mod}) can be obtained
by replacing the given solution by
$(\delta_l k_R(x) \delta_r, \delta_l \xi(x))$ with arbitrary $ \delta_l, \delta_r \in \bT_n$,
since (\ref{mod}) fixes the vector $\xi(x)$ up to phases,
according to (\ref{norx}),
 and the eigenvalues of
$e^{2\hat p}e^{\frac{x}{2}}$ are distinct.

\medskip

Let us rearrange the constraint (\ref{mod}) as
\be
k_R(x)^{-1}e^{2\hat p}e^{-\frac{x}{2}}k_R(x)=e^{2\hat p} e^{\frac{x}{2}}  -
2\left({\sinh}\frac{x}{2}\right) k_R(x)^{-1}\xi(x)\xi(x)^\dagger k_R(x).
\label{mod2}\ee
Let us also consider (\ref{mod}) for $x\to -x$,
\be
k_R(-x)e^{2\hat p}e^{-\frac{x}{2}}k_R(-x)^{-1}=e^{2\hat p} e^{\frac{x}{2}}  -
2\left({\sinh}\frac{x}{2}\right)\xi(-x)\xi(-x)^\dagger.
\label{mmod}\ee
By comparing (\ref{mod2}) and (\ref{mmod}), we conclude that
\be
(k_R(x)^{-1}\xi(x))_j =e^{\ri\eta_j} \xi(-x)_j,
\label{pm}\ee
where the $e^{\ri\eta_j} $ are some phases.
Indeed, this follows from the fact that the constraint (\ref{mmod}) determines the components
of  $\xi(-x)$ up to phases.

\medskip

Let us multiply the constraint (\ref{mod}) from the right by $k_R(x)$. This gives
 \be
 k_R(x)e^{2\hat p}e^{\frac{x}{2}} -e^{2\hat p} e^{-\frac{x}{2}}k_R(x)=
 2\left({\sinh}\frac{x}{2}\right)\xi(x)\xi(x)^\dagger k_R(x).
 \label{mod3}\ee
 Next, by  spelling out  (\ref{mod3}) in components and taking into account (\ref{pm}), we obtain
 \be
 \left(e^{2\hat p_l+\frac{x}{2}}-e^{2\hat p_j-\frac{x}{2}}\right) k_R(x)_{jl}=
 2\left({\sinh}\frac{x}{2}\right)\xi(x)_j\xi(-x)^\dagger_l e^{-\ri\eta_l}.
 \label{zab}\ee
For any $\hat p\in\C_x$, we can deduce from (\ref{zab}) that
 \be
 k_R(x)_{jl}= 2\left(\sinh \frac{x}{2}\right) \frac{e^{-\ri\eta_l}\xi(x)_j
 \xi(-x)^\dagger_l}{e^{2\hat p_l+\frac{x}{2}}-e^{2\hat p_j-\frac{x}{2}}}\,.
 \label{grs}\ee
  Now we note that if $\hat p\in\C_x$, then $\theta(x,\hat p)$,  defined by
(\ref{5.23a}) and (\ref{5.23b}),  can be also rewritten  as
\be
\theta(x,\hat p)_{jl}= 2  \left(\sinh\frac{x}{2}\right)
\frac{  \vert \xi(x)_j \vert\vert  \xi(-x)_l\vert}
{e^{2\hat p_l +\frac{x}{2}} - e^{2\hat p_j -\frac{x}{2}}}\,,
\label{thetaid1} \ee
 where  the absolute values of the components of the vectors  $\xi(\pm x)$
 are given by  (\ref{norx}).
It is clear from (\ref{grs}) and (\ref{thetaid1})  that that each unitary
solution $k_R(x)$ of (\ref{mod}) verifies
\be
k_R(x) = \delta_l \theta(x, \hat p) \delta_r
\quad\hbox{with some}\quad
\delta_l, \delta_r\in \bT_n,
\qquad
\forall \hat p\in \C_x,
\label{Laciform}\ee
which proves (\ref{tte}) for $\hat p\in \C_x$.
Moreover, the formula in (\ref{Laciform}) entails that the real matrix $\theta(x, \hat p)$ must be itself
unitary, and hence orthogonal,  and it must also
satisfy (\ref{mod})
for all $\hat p \in \C_x$.
By the continuity of $\theta(x,\hat p)$ as a function of $\hat p\in \bar \C_x$,
it then follows that $\theta(x, \hat p)$ must be an orthogonal matrix for all
$\hat p\in \bar\C_x$, and there must exists also a vector, say $\tilde \xi(x,\hat p)$, such that
\be
 \theta(x,\hat p)e^{2\hat p}e^{\frac{x}{2}}\theta(x,\hat p)^{-1}=
 e^{2\hat p}e^{-\frac{x}{2}} +2\left(\sinh\frac{x}{2}\right)\tilde \xi(x,\hat p)\tilde \xi(x,\hat p)^\dagger,
 \qquad \forall \hat p \in \bar\C_x.
 \label{mod30}\ee
Indeed, the fact that
$(\theta(x,\hat p)e^{2\hat p}e^{\frac{x}{2}}\theta(x,\hat p)^{-1}- e^{2\hat p}e^{-\frac{x}{2}})/(2\sinh\frac{x}{2})$
is a rank-one projector for all $\hat p\in \C_x$ implies that
the same statement holds also at the boundary of $\bar\C_x$.
(The rank cannot decrease at the boundary, since the vectors $\xi(x)$ satisfying (\ref{mod})
cannot vanish at any $\hat p\in \bar\C_x$.)
Thus we have shown
that $\theta(x,\hat p)$ is unitary and solves (\ref{mod}) for all $\hat p \in \bar \C_x$.
By the remarks given after (\ref{norx}),
this guarantees the validity of (\ref{tte})  at every $\hat p\in \bar \C_x$.
We note in passing that,
since the vector $\tilde \xi(x,\hat p)$ is determined by (\ref{mod30})
 up to an overall phase  at any fixed $\hat p$, and $\theta(x,\hat p)$ is real and continuous,
 the components of $\tilde \xi(x,\hat p)$ can be chosen to be real, continuous functions
 of $\hat p\in \bar\C_x$.
\emph{Q.E.D.}

\medskip
\noindent
{\bf Lemma 5.4.}
{\it If $K=k_L\trl(e^{-\hat p}k_R^{-1})$ (\ref{Carm}) is a solution
of the moment map constraint (\ref{MMC}) then the matrix $k_L$ must have the form
\be
k_L=h\kappa_L(x)\zeta(x,\hat p)^{-1}{\td}^{-1}.
\label{kle}\ee
Here $h\in G_{v(x)}$ (\ref{Gv}), ${\td}$ is some diagonal unitary matrix,
 the  matrix
$\kappa_L(x)$ is  given by (\ref{C}),  and   $\zeta(x,\hat p)$
is the real orthogonal  matrix  defined for every  $\hat p\in \bar \C_x$  by
\be
\zeta(x,\hat p)_{aa}= r(x,\hat p)_a, \  \zeta(x,\hat p)_{ij}=
\delta_{ij}-\frac{r(x,\hat p)_ir(x,\hat p)_j}{1+r(x,\hat p)_a},
\    \zeta(x,\hat p)_{ia}= - \zeta(x,\hat p)_{ai}= r(x,\hat p)_i, \quad i,j\neq a,
\label{CD}\ee
where $a=n$ for $x>0$, $a=1$ for $x<0$ and, for all $x\neq 0$,
\be
r(x,\hat p)_j:=\sqrt{\frac{1-e^{-x}}{1-e^{-nx}}}
\prod_{k\neq j}\sqrt{\frac{1-e^{2\hat p_j-2\hat p_k-x}}{1-e^{2\hat p_j-2\hat p_k}}},
\qquad
j=1,...,n.
\label{vew}\ee}

\medskip
\noindent
{\bf Proof.}
First,  let us show that
 the above real matrix $\zeta(x,\hat p)$ is orthogonal.
 For this, note that
 \be
 \sum_j\vert r(x,\hat p)_j\vert^2= 1.
 \label{jedn}\ee
 This can be deduced from the comparison of (\ref{vew}) and (\ref{eqw})
 by using  (\ref{char}) for $\lambda=0$.
One can then  easily check that
the columns of the matrix $\zeta(x,\hat p)$ form
an orthonormal system.

 We know from Eqs.~(\ref{defw}), (\ref{eqw}) and (\ref{vew}) that
 there exists ${\td}\in\bT_n$ such that
 \be
 k_L^{-1}v(x)=\sqrt{n}{\td} r(x,\hat p),
 \label{klv-extra}\ee
 where  the vector $r(x,\hat p)$ is defined by its components   (\ref{vew}).
 On the other hand, the formula (\ref{C}) leads immediately to
\be
\kappa_L(x)^{-1}v(x)=\left(\begin{matrix} 0\cr .\cr .\cr .\cr 0\cr \sqrt{n}\end{matrix}\right)
\quad\hbox{for}\,\,\,
x>0
\quad\hbox{and}\quad
 \kappa_L(x)^{-1}v(x)=\left(\begin{matrix} \sqrt{n}\cr 0\cr .\cr .\cr .\cr 0
 \end{matrix}\right)\quad\hbox{for}\,\,\, x<0.
 \label{matr}\ee
For all $x$,  we then obtain
 \be
 \zeta(x,\hat p)\kappa_L(x)^{-1}v(x)=\sqrt{n} r(x,\hat p).
 \label{mat}\ee
By combining this with (\ref{klv-extra}), we get
 \be
 {\td}^{-1}k_L^{-1}v(x)=\zeta(x,\hat p)\kappa_L(x)^{-1}v(x),
 \ee
 which implies the existence of an element $h$ from the isotropy group $G_{v(x)}$ of the
 vector $v(x)$ (\ref{vx}) such that (\ref{kle}) holds. \emph{Q.E.D.}

\medskip
\noindent
{\bf Lemma 5.5.}
{\it For each $\tau=\diag(\tau_1,\dots,\tau_n)\in\bT_n$  set
\be
\tau_{(x)}:=\diag(\tau_2,\dots,\tau_{n},1) \quad\hbox{if}\quad  x>0,  \qquad
\tau_{(x)}:=\diag(1,\tau_1,\dots,\tau_{n-1})\quad\hbox{if}\quad x<0.
\label{defoftx}\ee
Using the previous notations,
define the  map $K_x: G_{v(x)}\times \bT_n\times \bar\C_x \to \gln$
  by
 \be
 K_x(h,\tau,\hat p):=
 \left(h\kappa_L(x)\tau_{(x)}\zeta(x,\hat p)^{-1}\right)
 \trl \left(e^{-\hat p} \tau \tau_{(x)}^{-1}\theta(x,\hat p)^{-1} \right),
 \quad \forall (h,\tau,\hat p)\in G_{v(x)}\times \bT_n\times \bar\C_x.
 \label{kx}\ee
 Then the image of the map $K_x$ coincides with  the submanifold
 $F_{\nu(x)}=\Lambda^{-1}(\nu(x))$ of $\gln$.   }

\medskip
 \noindent {\bf Proof.}
  Consider   $h=\1_n$, $\tau=\1_n$ and $\hat p\in\bar\C_x$.
 We  first wish to show that $K_x(\1_n,\1_n,\hat p)$ solves the constraint (\ref{MMC}).
This statement is equivalent to
 \be
 \theta(x,\hat p)e^{2\hat p} e^{\frac{x}{2}}\theta(x,\hat p)^{-1}=
 e^{2\hat p} e^{-\frac{x}{2}} +2\left(\sinh\frac{x}{2}\right)\Xi(x,\hat p)\Xi(x,\hat p)^\dagger,
 \label{mod5}\ee
where the real vector $\Xi(x,\hat p)$ is defined as
 \be
 \Xi(x,\hat p):= \sqrt{\frac{e^{nx}-1}{e^x-1} }e^{\hat p}r(x,\hat p),
 \qquad
 \forall \hat p\in \bar\C_x,
 \label{Xi}\ee
 with $r(x, \hat p)$ in (\ref{vew}).
 This is so simply because (\ref{mat}) holds,
and the moment map constraint is (\ref{mod})
with (\ref{jv}).

Let us recall from the proof of Lemma 5.3  that
there exists a real  vector $\tilde \xi(x,\hat p)$ that satisfies
(\ref{mod30}).
We also know that $\tilde\xi(x,\hat p)$  verifies
 $\vert  \tilde\xi(x,\hat p)_j\vert = \Xi(x,\hat p)_j$  for all $j$,
 and is determined by (\ref{mod30}) up to an overall  sign.
  Notice from (\ref{norx}) that
  $\tilde\xi(x,\hat p)_n\neq 0$ holds for $x>0$ and  $\tilde\xi(x,\hat p)_1\neq 0$
  holds for $x<0$, at each $\hat p\in \bar \C_x$.
  This fact allows us
to fix the  sign ambiguity of $\tilde\xi(x,\hat p)$  by requiring that
$\tilde\xi(x,\hat p)_n=\Xi(x,\hat p)_n$ for $x>0$
and $\tilde\xi(x,\hat p)_1=\Xi(x,\hat p)_1$ for $x<0$.

 Now we are going to prove
  that the  unique vector $\tilde \xi(x,\hat p)$ specified above actually satisfies
  \be
   \tilde \xi(x,\hat p)=\Xi(x,\hat p), \qquad
   \forall \hat p\in \bar\C_x,
   \label{Lacimod}\ee
   which converts (\ref{mod30}) into (\ref{mod5}).
   We start by noting from Eqs.~(\ref{5.23a}) and (\ref{5.23b}) that
 \be
 \theta(-x,\hat p)=\theta(x,\hat p)^{-1}.
 \label{theta-extra}\ee
 Eqs.~(\ref{mod30}) and (\ref{theta-extra}) together imply
 \be
 \theta(-x,\hat p) e^{2\hat p}e^{-\frac{x}{2}}\theta(-x, \hat p)^{-1}
 =e^{2\hat p}e^{\frac{x}{2}} -2\left(\sinh \frac{x}{2}\right)\theta(x,\hat p)^{-1}\tilde\xi(x,\hat p)
 \tilde\xi(x,\hat p)^\dagger \theta(x,\hat p).
 \label{mod4}\ee
This entails
  \be
  \theta(x,\hat p)^{-1}\tilde\xi(x,\hat p) =\pm \tilde\xi(-x,\hat p),
  \label{pmm}\ee
  where the overall sign will be determined soon.
  Let us multiply Eq.~(\ref{mod30}) from the right by $\theta(x,\hat p)$. This gives
 \be
 \theta(x,\hat p)(x)e^{2\hat p}e^{\frac{x}{2}} -e^{2\hat p}e^{-\frac{x}{2}} \theta(x,\hat p)=
 2 \left(\sinh \frac{x}{2}\right) \tilde\xi(x,\hat p)\tilde\xi(x, \hat p)^\dagger  \theta(x,\hat p).
 \label{mod6}\ee
Using Eqs.~(\ref{pmm}) and (\ref{mod6}),  we then  find easily
   \be
   \theta(x,\hat p)_{jl}=
   \pm 2 \left(\sinh \frac{x}{2}\right)  \frac{\tilde\xi(x,\hat p)_j\tilde\xi(-x,\hat p)_l}{e^{2\hat p_l+\frac{x}{2}}-
   e^{2\hat p_j-\frac{x}{2}}}\,.
   \label{grsm}\ee
   On the other hand, we know from (\ref{thetaid1}) that
      \be
      \theta(x,\hat p)_{jl}= 2\left(\sinh\frac{x}{2}\right) \frac{\Xi(x,\hat p)_j\Xi(-x,\hat p)_l}{e^{2\hat p_l+
      \frac{x}{2}}-e^{2\hat p_j-\frac{x}{2}}}\,.
      \label{grsmm}\ee
      Let us restrict $\hat p$ to $\C_x$ in the above two equations
      (although both (\ref{grsm}) and (\ref{grsmm}) extend from
      $\C_x$ to $\bar \C_x$ by continuity).
      Comparing  Eqs.~(\ref{grsm}) and (\ref{grsmm}) for $j=n,l=1$ and  $j=1,l=n$
      fixes the positive sign
      in (\ref{pmm}) and (\ref{grsm}). Comparing  Eqs.~(\ref{grsm}) and (\ref{grsmm})
      for $j=n$, $l$ arbitrary
      and $l=1$, $j$ arbitrary
      gives then
 $\tilde\xi(x,\hat p)=\Xi(x,\hat p)$ for all $\hat p \in \C_x$.
 By the continuity of $\tilde \xi(x,\hat p)$ and $\Xi(x,\hat p)$ at every $\bar p\in\bar\C_x$, the
 claim of Eq.~(\ref{Lacimod}) then holds everywhere.

 \medskip

In the above we have proved that $K_x(\1_n,\1_n,\hat p)$ solves the moment map constraint
(\ref{MMC}).
Notice from (\ref{matr}) and (\ref{defoftx}) that
\be
\kappa_L(x)\tau_{(x)}\kappa_L(x)^{-1}\in G_{v(x)},
\qquad \forall \tau\in \bT_n.
\label{extratrick}\ee
It follows that whenever
$k_L$, $\hat p$ and $k_R$ solve   (\ref{Jrov}), then
 also $h\kappa_L(x)\tau_{(x)}\kappa_L(x)^{-1}k_L$, $\hat p$ and
  $ k_R \tau^{-1}\tau_{(x)}$ solve
 (\ref{Jrov}) for every $h\in G_{v(x)}$ and $\tau\in \bT_n$.
This means that also  $K_x(h,\tau,\hat p)$ solves the moment map constraint (\ref{MMC}).

 \medskip

Let us now show that all solutions of the moment map constraint (\ref{MMC})
are of the form $K_x(h,\tau,\hat p)$.
 Using the Lemmas  5.2,  5.3 and 5.4, we know that the most general solution of the moment map
 constraint must be of the form  $k_L\trl(e^{-\hat p}k_R^{-1})$, where $\hat p\in\bar\C_x$ and
 \be
 k_R=\tdl\theta(x,\hat p){\tdr}, \qquad
 k_L=h\kappa_L(x)\zeta(x,\hat p)^{-1}{\td}^{-1}
 \label{LR}\ee
 with arbitrary $h\in G_{v(x)}$ and certain $\tdl,{\tdr},{\td}\in \bT_n$.
 The substitution of (\ref{LR}) into the moment
 map constraint (\ref{Jrov}) gives
 \be
 \theta(x,\hat p)e^{2\hat p}\theta(x,\hat p)^{-1}=
 e^{2\hat p}e^{-x} +  e^{-x}\frac{e^{nx}-1}{n}e^{\hat p}\delta_l^{-1}{\delta}
 \zeta(x,\hat p)\kappa_L(x)^{-1} v(x)v(x)^{\dagger}\kappa_L(x)\zeta(x,\hat p)^{-1}{\td}^{-1}
 \tdl e^{\hat p}.
 \label{Jrov2}\ee
 On the other hand,   from Eqs.~(\ref{CD}), (\ref{matr}), (\ref{mod5}) and  (\ref{Xi}), we deduce
 \be
 \theta(x,\hat p)e^{2\hat p}\theta(x,\hat p)^{-1}=  e^{2\hat p}e^{-x} +
 e^{-x}
 \frac{e^{nx}-1}{n}e^{\hat p}   \zeta(x,\hat p)
 \kappa_L(x)^{-1} v(x)v(x)^{\dagger}\kappa_L(x)\zeta(x,\hat p)^{-1}
 e^{\hat p} .
 \label{Jrov3}\ee
 The compatibility of Eqs.~(\ref{Jrov2}) and (\ref{Jrov3}) requires that
 \be
 \delta_l^{-1} {\td} \zeta(x,\hat p) \kappa_L(x)^{-1} v(x) =
 \gamma \zeta(x,\hat p) \kappa_L(x)^{-1} v(x)
 \ee
with some phase $\gamma\in U(1)$. This then implies that
\be
\delta_l^{-1} {\td} \zeta(x,\hat p) \kappa_L(x)^{-1} = \gamma\zeta(x,\hat p) \kappa_L(x)^{-1} h'^{-1}
\ee
with some $h'\in G_{v(x)}$. By taking the inverse of this equation, we obtain
\be
\kappa_L(x) \zeta(x,\hat p)^{-1} {\td}^{-1} \tdl = h' \gamma^{-1} \kappa_L(x) \zeta(x,\hat p)^{-1}.
\label{propi}\ee
Finally, we conclude
\be
k_L\trl(e^{-\hat p}k_R^{-1})=\left[h\kappa_L(x)\zeta(x,\hat p)^{-1}\delta^{-1}\right]\trl
\left[e^{-\hat p}
\delta_r^{-1}\theta(x,\hat p)^{-1}\delta_l^{-1}\right]= K_x(h'',\aleph(x,\tdl {\tdr}), \hat p),
\label{required}\ee
where
\be
h''=hh'\kappa_L(x)\aleph(x,\tdl \tdr)_{(x)}^{-1}\kappa_L(x)^{-1} \in G_{v(x)}
\ee
with
\be
\aleph(x,\tau)_j:=\prod_{k=j}^n\tau^{-1}_k, \quad x>0
\quad\hbox{and}\quad
  \aleph(x,\tau)_j:=\prod_{k=1}^j\tau^{-1}_k, \quad x<0.
\label{deff}\ee
The proof is complete by the last equality in (\ref{required}).
To derive this, in addition to (\ref{propi}) we also employed the identity
\be
 \aleph(x,\tau)(\aleph(x,\tau))_{(x)}^{-1} = \tau^{-1},
 \qquad
 \forall \tau\in \bT_n,
 \label{alephprop}\ee
 which is satisfied by the bijection
 $\aleph(x,\cdot ): \bT_n \to \bT_n$ defined by (\ref{deff}).
\emph{Q.E.D.}
\medskip

      Recall  the following lemma   proved in \cite{Klong} (Lemma 5.1 therein).

\medskip
 \noindent
 {\bf  Lemma 5.6 (\cite{Klong}).} {\it  Let $A<\gln$ be the subgroup of
real
diagonal matrices with positive entries.
Consider the three maps $\iota, \iota_{L,R}: U(n)\times A \times U(n)\to \gln$
defined by
\be
\iota(\eta_L,a,\eta_R):=\eta_La\eta_R^{-1}, \quad
  \iota_L(\eta_L,a,\eta_R):=\eta_La,\quad
  \iota_R(\eta_L,a,\eta_R):=a\eta_R^{-1}.
  \label{iotadef}\ee
  Then  the $\iota$-pullback  and $ \iota_{L,R}$-pullbacks of the
  symplectic form $\om_+$   given by (\ref{ST})
 are related as
\be
\iota^*\om_+=\iota_L^*\om_++\iota_R^*\om_+.
\ee}

 \noindent {\bf  Lemma 5.7.} {\it Using the notations (\ref{kx}) and (\ref{deff}),
  define the smooth map $k_x$ from the phase space $\hat P= \bT_n \times \C_x$ of the dual
  trigonometric Ruijsenaars-Schneider system into $\gln$ by
 \be
 k_x(\hat q,\hat p):=K_x(\1_n,\aleph(x,e^{\ri\hat q}),\hat p).
 \label{dmr}\ee
 Then it holds
 \be
 k^*_x\om_+=\hat \om,
 \label{kxpull}\ee
 where $\hat \om$ is the symplectic form on $\hat P$ given by (\ref{1.4}).}

 \medskip
 \noindent {\bf Proof.}
  Define the map  $\eta: \bT_n\times \C_x\to U(n)\times A\times U(n)$
 by
 \be
 \eta(\tau,\hat p):=(\eta_L(\hat p), e^{-\hat p},  \eta_R(\tau,\hat p)),
 \qquad \forall \tau\in\bT_n,\quad \forall \hat  p\in \C_x,
 \ee
 where
 \be
 \eta_L(\hat p):=\kappa_L(x)\zeta(x,\hat p)^{-1}, \qquad \eta_R(\tau,\hat p):=
 \Xi_R(\kappa_L(x)\zeta(x,\hat p)^{-1}e^{-\hat p})^{-1}
 \theta(x,\hat p)\tau_{(x)}\tau^{-1}.
 \ee
It follows from Eqs.~(\ref{2.21}), (\ref{kx}) and (\ref{iotadef})
  that
  \be
  K_x(\1_n,\tau,\hat p)=(\kappa_L(x)\tau_{(x)}\kappa_L(x)^{-1})\trl\iota (\eta(\tau,\hat p)).
  \label{jot}\ee
 Now  we wish to calculate the pull-back
  $K_x(\1_n)^*\om_+$ where the map $K_x(\1_n):\bT_n\times \C_x\to \gln$ is
  defined by $K_x(\1_n)(\tau,\hat p):=K_x(\1_n,\tau,\hat p)$.
  The factor $\kappa_L(x)\tau_{(x)}\kappa_L(x)^{-1}$   does not contribute to this pull-back,
 since $K_x(\1_n)$ takes its values in $F_{\nu(x)}$ by Lemma 4.5 and
$\kappa_L(x)\tau_{(x)}\kappa_L(x)^{-1}$ acts by  gauge transformations
 on $F_{\nu(x)}$
 because of (\ref{extratrick}).
 By combining (\ref{jot}) with Lemma 4.6, we then obtain
  \be
  K_x(\1_n)^*\om_+=\eta^*\iota^*\om_+ = \eta^*\iota^*_L\om_+ + \eta^*\iota_R^*\om_+.
  \ee
 Next, notice that
  \be
  \eta^*\iota_L^*\om_+=0.
  \ee
This is because  the element $\eta_L(\hat p)e^{-\hat p}$ is real, the
images of the Iwasawa maps $\Lambda_L,\Lambda_R,\Xi_R,\Xi_L$ are
therefore real, too, and hence the imaginary part of the expression
under the trace in  (\ref{ST}) vanishes in this case.

\medskip

 In order to find   $ \eta^*\iota^*\om_+ = \eta^*\iota_R^*\om_+$, we   have to calculate
\be
\Lambda_L(a\eta_R^{-1})=a,\quad \Xi_R(a\eta_R^{-1})=\rho \tau_{(x)} \tau^{-1}, \quad
\Xi_L(a\eta_R^{-1})=\tau \tau_{(x)}^{-1}\mu,\quad \Lambda_R(a\eta_R^{-1})=\rho a^{-1}\mu,
\ee
where
\be
\rho:= \eta_R(\tau,\hat p)\tau \tau_{(x)}^{-1},\qquad \mu:=\Xi_L(e^{-\hat p}\rho^{-1}).
\ee
By using the  formula (\ref{ST}), we compute directly
 \bea
&&\phantom{XXXXX}  \eta^*\iota_R^*\om_+=
 \jp\Im\tr\left(d\hat p\w (d\tau_{(x)} \tau_{(x)}^{-1}-d\tau \tau^{-1}-
 \tau \tau_{(x)}^{-1}d\mu\mu^{-1} \tau_{(x)}\tau^{-1})\right)+
 \phantom{XXX}\nonumber\\
 && +\jp\Im\tr\left((d\rho\rho^{-1} +\rho d\hat p\rho^{-1}
 +\rho e^{\hat p}d\mu\mu^{-1}e^{-\hat p}\rho^{-1})\w (d\rho\rho^{-1} +
 \rho (d\tau_{(x)}\tau_{(x)}^{-1}-d\tau \tau^{-1}) \rho^{-1})\right).
 \phantom{XXX}
 \eea
 By eliminating the terms that vanish due to the fact that the
 imaginary part of a real number vanishes, it remains
 \be
 \eta^*\iota_R^*\om_+=
 \jp\Im\tr \left((2d\hat p+\rho^{-1}d\rho+d\mu\mu^{-1})\w
 (d \tau_{(x)} \tau_{(x)}^{-1}-  d\tau \tau^{-1})\right).
 \ee
 Finally, we note that both $\rho$ and $\mu$ are real orthogonal matrices, hence both $\rho^{-1}d\rho$ and
 $d\mu\mu^{-1}$  are $1$-forms taking values in the space of antisymmetric matrices. Since
 the trace of the product
 of an antisymmetric matrix with a diagonal matrix vanishes, we arrive at
  \be
  K_x(\1_n)^*\om_+= \eta^*\iota_R^*\om_+
 = \Im\tr \left(d\hat p\w \bigl(\tau \tau_{(x)}^{-1} d (\tau_{(x)} \tau^{-1})\bigr)\right) .
 \label{jmr}\ee
 With the help of the relation
 $k_x(\hat q,\hat p) =K_x(\1_n)(\aleph(x,e^{\ri\hat q}),\hat p)$ and
 (\ref{alephprop}),
 we conclude from (\ref{jmr}) that
 \be
 k^{*}_x\om_+=\sum_{j} d\hat p_j \wedge d\hat q_j = \hat \om.
 \ee
 \emph{Q.E.D}

   \medskip\medskip
 \noindent {\bf Proof of Theorem 3.3:}
 First of all we remind that the dual Ruijsenaars-Schneider phase space $\hat P$ is identical to
  $\bT_n\times\C_x$ as a manifold (cf.~Remark 3.4).  The map $\Z_x:\hat P\to \hat P_c$,
 viewed as the map $\Z_x:\bT_n\times \C_x\to \hat P_c$, can be extended to  a {\it surjective}
 map $\tilde\Z_x:\bT_n\times \bar\C_x\to\hat P_c$ by using the
  same formulae (\ref{x+}) and (\ref{x-}) that define $\Z_x$.
  In correspondence with the Cartesian product $\hat P_c = \bC^{n-1} \times \bC^\times$,
we denote the components of this smooth map $\tilde \Z_x$ by $\tilde z$
and $\tilde Z$.
Then it is  straightforward  to check that the following identities hold for all
 $(e^{\ri \hat q}, \hat p)\in \bT_n \times \bar \C_x$:
\be
\hat\zeta(x,\tilde z(x, \hat q,\hat p))= \tau_{(x)}\zeta(x,\hat p)\tau_{(x)}^{-1},
\ee
\be
\hat\theta(x,\tilde z(x, \hat q,\hat p))= \tau_{(x)}\theta(x,\hat p)\tilde \tau_{(x)}^{-1},\ee
\be \Delta(x, \tilde z(x,\hat q,\hat p), \tilde Z(x,\hat q,\hat p))=
\tau\tilde \tau_{(x)}^{-1} e^{-\hat p},
\ee
where $\tau=\aleph(x,e^{\ri\hat q})$ as in (\ref{deff}), $\tau_{(x)}$ is given by (\ref{defoftx}), and
for every $\tau\in \bT_n$ we employ
\be
\tilde \tau_{(x)}:=\diag(1,\tau_2,...,\tau_n) \quad\hbox{if}\quad x>0,
\qquad
 \tilde \tau_{(x)}:=\diag(\tau_1,...,\tau_{n-1},1)\quad\hbox{if}\quad x<0.
\label{tildetx}\ee
 Recall that the orthogonal  matrices $\theta(x,\hat p)$, $\zeta(x,\hat p)$ were defined
in Lemma 5.3 and 5.4, respectively, and the matrices  $\hat\zeta(x,z)$ and
 $\hat\theta(x,z)$ were introduced in Definition 3.3. The surjectivity of the map
 $\tilde\Z_x:\bT_n\times \bar\C_x \to \bC^{n-1}\times \bC^\times$
 now implies the unitarity of the matrices $\hat\zeta(x,z)$ and $\hat\theta(x,z)$
 for every $x\neq 0$ and $z\in \bC^{n-1}$.

The above identities give rise to the relation
\be
(\hat \I\circ \tilde\Z_x)(\hat q,\hat p)=
K_x(\1_n,\aleph(x,e^{\ri\hat q}),\hat p), \qquad \forall e^{\ri\hat q}\in \bT_n,
\quad \forall \hat p\in \bar\C_x,
\label{bla}\ee
where the map $\hat \I$ was defined in Eq.~(\ref{dif}) and $K_x$ in Eq.~(\ref{kx}).
This implies by Lemma 5.5 and the surjectivity of
the maps $\tilde \Z_x$, $\aleph(x,.)$
that every point of the constraint-manifold
$F_{\nu(x)}=\Lambda^{-1}(\nu(x))$ can  be written as
$h\trl \hat \I(z,Z)$ with some $h\in G_{v(x)}$ and some
$(z,Z)\in \bC^{n-1}\times \bC^\times$.
In particular, the image of $\hat\I$ intersects every gauge orbit in $F_{\nu(x)}$.

Next, we wish to show that
\be
\hat \I^*\om_+=\ho.
\label{fr}\ee
Referring to (\ref{dmr}),
the restriction of the relation (\ref{bla}) to $\bT_n\times \C_x$ can be expressed as
\be
k_x = \hat\I \circ \Z_x.
\label{kxid}\ee
By Eqs.~(\ref{SRembed}) and (\ref{kxpull}), we have
   \be
   \Z_x^*\ho=\hat \omega=   k^{*}_x\om_+= \Z_x^*\hat \I^*\om_+.\label{gle}
   \ee
   We recall from Remark 3.5 that
    the map  $\Z_x$ yields  a diffeomorphism between $\hat P$ and the
dense open submanifold  $\hp^0$ of $\hp$  given by
  \be
  \hp^0:=\{(z,Z)\in \bC^{n-1}\times \bC^\times\,\vert\,  z_j\neq 0,\,\forall   j=1,...,n-1\}.
  \label{4.89}\ee
   We  thus see from Eq.~(\ref{gle})  that the pull-back  $\hat \I^*\om_+$  and  the form $\ho$
   coincide   everywhere on
   the dense open subset $\hp^0\subset \hp$.  Therefore, by smoothness, they coincide everywhere
   on $\hp$.

 \medskip

  For $h\in G_{v(x)}$ and $(z,Z),(z',Z')\in\hat P_c$, we now  prove the following implication:
 \be
h\trl \hat \I(z,Z)=\hat \I(z',Z')\implies h=\1_n, \quad z=z', \quad Z=Z'.
\label{impl} \ee
We present the argument in detail for the case $x>0$
 and leave to the reader the analogous case $x<0$. The assumption on the left hand side of (\ref{impl})
 entails that   $h\trl  \hat \I(z,Z)$ and  $\hat \I(z',Z')$  have the same matrix $e^{-\hat p}$
in the modified Cartan decomposition (\ref{Carm}).
By means of the formulae (\ref{D}), (\ref{D'}) and (\ref{dif}), this gives
\be
\Delta(x,z, \vert Z\vert )=\Delta(x,z',\vert Z'\vert ),
\ee
whence
\be
\vert Z'\vert =\vert Z\vert, \qquad \vert z'_i\vert =\vert z_i\vert , \quad \forall i=1,...,n-1.
\label{abs}\ee
Define the diagonal matrix $\Upsilon=\diag(\Upsilon_1,\dots, \Upsilon_{n-1},1)$ by
setting $\up_j :=\frac{z'_j}{z_j}$
if $\vert z_j\vert =\vert z'_j\vert \neq 0$ and $\up_j:=1$ otherwise.   The inspection
of the formulae   (\ref{sfc}), (\ref{hth1}) and  (\ref{hth2}) together with Eq.~(\ref{abs}) leads
to the following equalities:
\be
\up \hat\zeta(x,z)\up^{-1}= \hat\zeta(x,z'), \qquad
\up \hat\theta(x,z)\hat\up^{-1}= \hat\theta(x,z'),\label{prv}\ee
where $\hat\up:=\diag(1,\up_1,...,\up_{n-1})$.
Furthermore, the equality
\be
(h\trl \hat \I(z,Z))(h\trl \hat \I(z,Z))^\dagger=\hat \I(z',Z') \hat \I(z',Z')^\dagger
\ee
implies
\be
h \kappa_L(x)\hat\zeta(x,z)^{-1}\Delta(x,z,\vert Z\vert )^2\hat\zeta(x,z)\kappa_L(x)^{-1}h^{-1}=
\kappa_L(x)\hat\zeta(x,z')^{-1}\Delta(x,z,\vert Z\vert )^2\hat\zeta(x,z')\kappa_L(x)^{-1}.
\ee
This gives
\be
\hat\zeta(x,z)\kappa_L(x)^{-1} h^{-1}\kappa_L(x)=\Gamma \hat\zeta(x,z'),\label{dru}
\ee
for some $\Gamma=\diag(\Gamma_1,\ldots, \Gamma_n)\in \bT_n$ and, consequently,
\be
\hat\zeta(x,z)\kappa_L(x)^{-1}v(x) =\hat\zeta(x,z)
\kappa_L(x)^{-1} h^{-1}\kappa_L(x)\kappa_L(x)^{-1}v(x)=\Gamma \hat\zeta(x,z')\kappa_L(x)^{-1}v(x).
\label{499}\ee
By using  Eqs.~(\ref{matr}),   (\ref{aca}) and  the first equation of (\ref{prv}),
we find from (\ref{499}) that
\be
\Gamma_n=1
\quad\hbox{and}\quad
\Gamma_j^{-1}=\up_j \quad
\forall j\in\{1,...,n-1\}\quad\hbox{ for which}\quad z_j\neq 0.
\label{500}\ee
By multiplying Eq.~(\ref{dru}) by $\Gamma^{-1}$ from the right and by using the first
equation of (\ref{prv}), we then arrive at
\be
\hat\zeta(x,z)\kappa_L(x)^{-1} h^{-1}\kappa_L(x)\Gamma^{-1}=  \hat\zeta(x,z),\label{drub}
\ee
and thus
\be
\kappa_L(x)^{-1}h\kappa_L(x)=\Gamma^{-1}.
\label{prt}\ee
Now we return  to the equation  on the left hand side of the implication (\ref{impl})  and,
by using Eqs.~(\ref{dif}), (\ref{prv}) and (\ref{prt}), we rewrite
it as
$$
\left[h\kappa_L(x)\hat\zeta(x,z)^{-1}\right]\trl \left[\Delta(x,z,Z)\hat\theta(x,z)^{-1}\right]=
 \left[\kappa_L(x)\Gamma^{-1}\hat\zeta(x,z)^{-1}\right]\trl \left[\Delta(x,z,Z)\hat\theta(x,z)^{-1}\right]=
 $$
 \be
  \left[\kappa_L(x)\Gamma^{-1}\hat\zeta(x,z)^{-1}\Gamma\right]
 \trl \left[\Delta(x,z,Z)\Gamma^{-1}\hat\theta(x,z)^{-1}\Gamma\right]
 =
\left[\kappa_L(x)\hat\zeta(x,z')^{-1}\right]\trl
 \left[\Delta(x,z',Z') \hat\theta(x,z')^{-1} \right].
 \label{long}\ee
 We remark that $\Gamma^{-1}\hat\zeta(x,z)^{-1}\Gamma=
 \up\hat\zeta(x,z)^{-1}\up^{-1}=\hat\zeta(x,z')^{-1}$, and hence   from
 Eq.~(\ref{long})  we
 obtain
\be
\Delta(x,z',Z')\hat\theta(x,z')^{-1} =
 \Delta(x,z,Z)\Gamma^{-1}\hat\theta(x,z)^{-1}\Gamma.
 \ee
 Taking into account Eq.~(\ref{prv}), this yields
 \be
 \up\hat\theta(x,z)\hat\up^{-1} \Delta(x,z',Z')^{-1}=
 \Gamma^{-1}\hat\theta(x,z)\Gamma  \Delta(x,z,Z)^{-1}.
 \label{klu}\ee
 The $(j,j+1)$  entries of the matrices on the two sides of Eq.~(\ref{klu})
 never vanish.
 Together with (\ref{500}), the equality of these entries
 implies  that $\Gamma=\1_n= \up^{-1}$.
 We conclude that $h=\1_n$ by (\ref{prt}) and $z'=z$ by the definition of $\up$.
 Similarly, the $(n,1)$ entry never vanishes in (\ref{klu}), and this gives
 $Z'=Z$. The implication (\ref{impl}) is therefore proven.

\medskip
It is clear from its formula (\ref{dif}) and Lemma 3.2 that
$\hat \I: \hp \to F_{\nu(x)}$ is a \emph{smooth} map.
We see from the implication  (\ref{impl})   that the map $\hat\I$ is injective and
its image intersects every gauge orbit at most in one point.
Since we have shown also that the image of $\hat \I$ intersects every gauge orbit,
we conclude that condition 2) of Lemma 3.1 is satisfied.
Eq.~(\ref{fr}) guarantees  that condition 1) of the same lemma holds.
In conclusion,
 $\hat\I$ is a global cross section and
$(\hp, \hat\omega_c)$ is a model of the reduced phase space.
\emph{Q.E.D.}

\medskip
\medskip
\noindent {\bf Proof of Theorem 3.4:}
 We first observe from (\ref{bla}) and the definition of $\hat \cL$ (\ref{Unlax}) that
 \be
 (\hat{\cal L}\circ \hat\I\circ\Z_x)(\hat q,\hat p)=
 \hat {\cal L}(K_x(\1_n,\aleph(x,e^{\ri\hat q}),\hat p))=
 \Xi_R(K_x(\1_n,\aleph(x,e^{\ri\hat q}),\hat p)), \quad \forall(\hat q,\hat p)\in \hat P.
 \ee
 Using the notation $X\sim Y$ to signify that the matrices $X$ and $Y$ are similar,
 we then conclude from  (\ref{2.33}) and the formula (\ref{kx}) of $K_x$  that
 \be
 \Xi_R(K_x(\1_n,\aleph(x,e^{\ri\hat q}),\hat p))
 \sim \Xi_R(e^{-\hat p} (\aleph(x,e^{\ri\hat q}))_{(x)}^{-1}
 \aleph(x,e^{\ri\hat q}) \theta(x, \hat p)^{-1})
 =\theta(x,\hat p)e^{\ri\hat q}.
 \label{sim}\ee
 To obtain the last equality, we used the identity
 (\ref{alephprop}).
 The similarity transformation
 that appears in (\ref{sim}) is by a unitary matrix that one can find  explicitly
 from the above.
 On the other hand,
 one can directly check
 with the aid of the formula of  $\theta(x, \hat p)$ given in Lemma 5.3 that
 the dual Ruijsenaars-Schneider Lax matrix (\ref{1.3})
  can be   rewritten as
 \be
 \hat L(\hat q,\hat p)=
  \hat\delta  \theta(x,\hat p) e^{\ri\hat q} {\hat\delta}^{-1}
 \label{hatLconj}\ee
 with
 \be
 \hat\delta=\diag(\hat\delta_1,\hat\delta_2,\dots,\hat\delta_n), \quad
 \hat \delta_j=\prod_{m\neq j}
 \biggl[\frac{\sinh{(\hat p_j-\hat p_m+\fp)}}{\sinh{(\hat p_j-\hat p_m-\fp)}}\biggr]^{\frac{1}{4}},
 \quad \forall j=1,...,n.
 \ee
 Therefore the matrices $(\hat{\cal L}\circ \hat\I\circ\Z_x)(\hat q,\hat p)$ and
 $ \hat L(\hat q,\hat p)$ are both similar to the matrix
$\theta(x,\hat p)e^{\ri\hat q}$.  \emph{Q.E.D.}

\medskip
The implication (\ref{impl}) also confirms that $G_{v(x)}$ acts freely on
$F_{\nu(x)}$, as was already shown in \cite{LMP}.
Here, some further clarifying remarks are in order, which will be referred to
 in Section 6.

\medskip
\noindent
{\bf Remark 5.1.}
Let $F_{\nu(x)}^0$ be the image of $G_{v(x)} \times \bT_n \times \C_x$ by the
map $K_x$ (\ref{kx}). It follows from the above that $F_{\nu(x)}^0$ is a dense open
submanifold of $F_{\nu(x)}$, which is
stable under the gauge group $G_{v(x)}$ and is diffeomorphic
to $G_{v(x)} \times \bT_n \times \C_x$  by $K_x$.
Hence $F_{\nu(x)}^0/G_{v(x)}$ is a dense  open  submanifold of the full reduced
phase space $F_{\nu(x)}/G_{v(x)}$.
The phase space
$(\hat P, \hat \omega)$ (\ref{1.4}) is a model
of this dense open submanifold.
In fact,   a corresponding
cross section (in the sense of Remark 3.2) is provided by that map  $k_x:\hat P \to F_{\nu(x)}^0$ (\ref{dmr})
that
satisfies $k_x = \hat\I \circ \Z_x$ (\ref{kxid}) with the symplectic embedding $\Z_x:\hat P\to \hat P_c$
given by Definition 3.2.

\medskip
\noindent
{\bf Remark 5.2.}
We can define the $\bR^n$-valued smooth (even real-analytic) $G_{v(x)}$-invariant function
$\hat \pi$ on $F_{\nu(x)}$ by setting $\hat \pi  (K_x(h, \tau, \hat p)) := \hat p$.
More directly,  $\hat \pi\simeq \diag(\hat\pi_1,\ldots,\hat\pi_n)$ can be characterized by the property
\be
K K^\dagger \sim e^{- 2\hat\pi(K)} \quad\hbox{with}\quad \hat \pi(K) \in \bar \C_0,
\label{pihat}\ee
where, as before, $\sim $  denotes   similarity of matrices.
Clearly,   $\hat \pi$
induces a smooth function on the full reduced phase space and
it has $\bar \C_x$ as its range.
 From the perspective
of the completed dual Ruijsenaars-Schneider system, $\hat \pi$  can be viewed
as a $\bar \C_x$-valued globally well-defined `position variable'
(which coincides with $\hat p$ on the phase space $\hat P$).

\medskip
We explained in \cite{LMP} how Theorems 3.1 and 3.2 imply
the known integration algorithm for the time development of the
position variable $q$ along the commuting flows of the system $(P,\omega, L)$.
Now we present the analogous result about the `dual position variable' $\hat \pi$ (\ref{pihat})
along the commuting flows of the dual system
characterized by Theorems 3.3 and 3.4.

\medskip
\noindent
{\bf Remark 5.3.}
Take a Hamiltonian $\hat H:= \Xi_R^* \phi$, $\phi \in C^\infty(G)^G$
and regard it as a function on the model  $(\hat P_c, \hat\omega_c)$ of the reduced phase space.
Choose also an initial value $(z(0), Z(0))\in \hat P_c$.
Directly from the definitions, the associated initial value
$\hat\pi(0):= \hat \pi(\hat \I(z(0), Z(0)))$ is subject to
\be
e^{- \hat \pi(0)} = \Lambda_L(\Delta_0)
\quad\hbox{with}\quad
\Delta_0:=\Delta(x, z(0), Z(0)).
\label{I1}\ee
By combining Theorem 3.3 with Corollary 2.1,   it is easy to obtain the following result:
\be
e^{-2 \hat \pi(t)} \sim e^{-2 \hat \pi(0)}
\exp\bigl[2\ri t \bD\phi(\hat \theta(x, z(0)) \Xi_R(\Delta_0))\bigr].
\label{I2}\ee
Since  (\ref{pihat}) defines $\hat \pi$ also as a $\bar G$-invariant function on  the double,
(\ref{I2}) follows immediately from (\ref{2.53})  by replacing the initial value
$\hat \I(z(0), Z(0))$ (\ref{dif}) by $\Delta(x, z(0), Z(0)) \hat\theta(x, z(0))^{-1}$.
Indeed, the respective solutions $K(t)$ (\ref{2.48}) are $\bar G$-related and yield the same $\hat \pi(t)$.

If $(z(0), Z(0))\in (\hat P_c \setminus \hat P_c^0)$ (see (\ref{4.89})), then
the flow (\ref{I2}) does not stay in this lower dimensional closed submanifold in general, which explains
by Theorem 3.4
why the commuting dual flows are not complete on $(\hat P, \hat \omega)$.
On the other hand,
if $(z(0), Z(0))\in \hat P_c^0$ and $(\hat q(0), \hat p(0))$ is
the corresponding initial value in $\hat P= \Z_x^{-1}(\hat P_c^0)$, then
(\ref{I2}) can be rewritten as
\be
e^{- 2\hat \pi(t)} \sim e^{-2 \hat p(0)}
\exp\bigl[2\ri t \bD\phi(\theta(x,\hat p(0))e^{\ri \hat q(0)})\bigr].
\label{I3}\ee
To further elaborate, let us consider $\phi(g) = \tr(\chi(g)) + \mathrm{c.c.}$
with some complex power series $\chi$, for which
$\bD\phi(g) = \psi(g) - (\psi(g))^\dagger$ with $\psi(z):= z \chi'(z)$ as was mentioned
in (\ref{2.58}).
For $\psi(z) = \sum_{k=1}^\infty \psi_k z^k$, let us define
$\tilde \psi(z):= \sum_{k=1}^\infty \bar\psi_k z^k$, where $\bar \psi_k$ is the complex conjugate
of $\psi_k$.
Then, by using (\ref{hatLconj}), we can rewrite ({\ref{I3})
equivalently as
\be
e^{- 2\hat \pi(t)} \sim e^{- 2\hat p(0)}
\exp\bigl[2\ri t ( \psi(\hat L_0) - \tilde \psi({\hat L_0}^{-1}))\bigr]
\quad
\hbox{with}\quad
\hat L_0:= \hat L(\hat q(0), \hat p(0)).
\label{I4}\ee
These formulas for the flows were obtained previously
in \cite{RIMS95}.
Our geometric picture renders their derivation essentially obvious.

\section{  Discussion}
\setcounter{equation}{0}

In this section, we summarize
our construction in terms of diagrams of maps and explain the connection
with the related results in
\cite{RIMS95}. Then we conclude and comment  on open problems.

Our main arguments were concerned with the following diagram:
\be
\begin{array}{ccc}
P & \stackrel{\I }{\longrightarrow}  & F_{\nu(x)}/G_{v(x)} \\
{} & {} & {}\\
 {\scriptstyle{\cR }} \, \downarrow   \phantom{X}  & {} &  \phantom{X}   \downarrow \,   {\scriptstyle{\mathrm{id}}} \\
{} & {} & {} \\
\hat P_c & \ \stackrel{\pi \circ \hat{\I}} \longrightarrow   &  F_{\nu(x)}/G_{v(x)} \\
\end{array}\label{diagram1}
\ee
The maps $ \I$ and $\pi \circ \hat \I$ are symplectomorphisms as claimed by
Theorems 3.1 and 3.3.  The map  $\cR$ is defined as
the symplectomorphism that makes this diagram commute.
We can add the phase space $(\hat P, \hat \omega)$ (\ref{1.4}) to (\ref{diagram1}) by using
the  embedding $\Z_x$ of Definition 3.2 and restrictions to
the relevant dense open submanifolds.
This leads to the second diagram:
\be
\begin{array}{ccccc}
P^0 &  \stackrel{\mathrm{id}}{\longrightarrow} &  P^0 & \stackrel{{\I }^0}{\longrightarrow}
& F^0_{\nu(x)}/G_{v(x)} \\
{}& {} & {} & {}& {}  \\
{\scriptstyle{\cR^0}} \, \downarrow  & \phantom{X} & {\scriptstyle{{{\cR }^0_c}}} \,
\downarrow  & \phantom{X} &   \downarrow \, {\scriptstyle{{\mathrm{id}}}} \\
{}& {} & {} & {}& {}  \\
\hat P &  \stackrel{\Z_x}{\longrightarrow} &   \hat P^0_c
& \ \stackrel{\pi\circ {\hat{\I}^0}} \longrightarrow  & F^0_{\nu(x)}/G_{v(x)} \\
\end{array}\label{diagram2}
\ee
Here, $P^0 \subset P$ and $\hat P_c^0 \subset \hat P_c$ are the
dense open submanifolds symplectomorphic to the dense open submanifold
$F_{\nu(x)}^0/G_{v(x)}$ of the reduced phase space, for which $\hat \pi$
(\ref{pihat}) varies in $\C_x$ (\ref{gap}).
(See also (\ref{4.89}) and Remark 5.1.)
Whenever
appropriate, the maps in (\ref{diagram2}) are obtained as the restrictions of those in (\ref{diagram1}).
The  map $\cR^0$ is defined by the commutativity of the diagram (\ref{diagram2}).
The symplectic forms $\omega$ (\ref{1.2}) on $P$, $\hat \omega_c$ (\ref{hatom}) on $\hat P_c$,
and the reduced symplectic form on $F_{\nu(x)}/G_{v(x)}$ induce symplectic forms
on the respective dense open submanifolds. Then all the maps
$ \I^0$, $\pi \circ \hat \I^0$,  $\cR_c^0$, $\Z_x$ and $\cR^0$
 are symplectomorphisms.

The maps $\cR$ and $\cR^0$ reproduce
the `complete' and the `restricted' duality symplectomorphisms (alias action-angle maps) originally obtained  by
Ruijsenaars in \cite{RIMS95}  by means of  direct arguments.
In order to confirm this, we now present a useful consequence of our
construction.

\medskip
\noindent
{\bf Lemma 6.1.}
\emph{
Consider a point $(q,p)\in P^0$ and its image  $(\hat q, \hat p)\in \hat P$
by the symplectomorphism  $\cR^0$ defined by (\ref{diagram2}):
\be
\cR^0(q,p) = (\hat q, \hat p),
\label{D4}\ee
where $(q,p)$ parametrizes some point $(e^{2\ri q}, p)$ of the covering space
$T^* \bT_n^0$ (\ref{3.17}) of $P$ (\ref{3.16}).
Then the Lax matrices $L(q,p)$ (\ref{1.1}) and $\hat L(\hat q, \hat p)$  (\ref{1.3}) satisfy
the similarity relations
\be
L(q,p) \sim e^{2 \hat p}
\quad\hbox{and}\quad
\hat L(\hat q, \hat p) \sim e^{2\ri q}.
\label{D5}\ee}

\medskip
\noindent
{\bf Proof.}
First,
let us recall from Theorem 3.2 that
$L(q,p) \sim (\cL \circ \tilde \I)(e^{2\ri q}, p)$.
The definition (\ref{Unlax})
of the unreduced Lax matrix $\cL$ gives
$(\cL\circ \tilde \I)(e^{2\ri q}, p) =
\tilde \I(e^{2\ri  q},  p)^{-1} (\tilde \I(e^{2\ri  q}, p)^{-1})^\dagger$,
and then we obtain
\be
\tilde\I(e^{2\ri q}, p)^{-1} (\tilde\I(e^{2\ri  q}, p)^{-1})^\dagger \sim
(\tilde \I(e^{2\ri  q}, p) \tilde\I(e^{2\ri q}, p)^\dagger)^{-1}
\sim \bigl(((\hat \I\circ \Z_x)(\hat q,\hat p)) ((\hat\I\circ \Z_x)(\hat q,\hat p))^\dagger\bigr)^{-1}
\sim e^{2\hat p}.
\label{D6}\ee
Here we used that, because of  ({\ref{D4}),
 $\tilde \I(e^{2\ri  q}, p)$ and $(\hat \I\circ \Z_x)(\hat q, \hat p)$
must lie on the same gauge orbit in $F_{\nu(x)}^0$,
which implies the second similarity relation in (\ref{D6}) on account of the first relation
in (\ref{2.33}).
The last relation in (\ref{D6}) follows from (\ref{pihat}), since
$\hat \pi(\hat\I\circ \Z_x(\hat q, \hat p)) = \hat p$ holds on  $\hat P$.
Thus the first claim of (\ref{D5}) is proved.
To prove the second claim of (\ref{D5}), we notice from Theorem 3.4
together with the definition of $\hat \cL$ in
(\ref{Unlax})
and the second relation in (\ref{2.33}) that
\be
\hat L(\hat q, \hat p) \sim (\hat \cL \circ \hat \I \circ \Z_x)(\hat q, \hat p) \sim
\Xi_R( \tilde \I(e^{2\ri q},p))=e^{2 \ri q}.
\ee
The last equality above is a direct consequence of the formula (\ref{N.19}) of
$\tilde \I(e^{2\ri q},p)$. \emph{Q.E.D.}
\medskip

We see from (\ref{D5}) that our symplectomorphism $\cR^0$ between the restricted phase spaces
$P^0$ and $\hat P$ converts the action variables of the original system (\ref{1.1})
into the particle-coordinates of the dual system (\ref{1.3}), and vice versa.
These relations completely characterize the map $\cR^0$ and also hold for the
restricted duality symplectomorphism constructed in \cite{RIMS95}.
Therefore the latter is indeed
reproduced by our geometric construction.
 The complete duality symplectomorphism $\cR$ of diagram (\ref{diagram1}) is also the
 same
as the one obtained in \cite{RIMS95}\footnote{To provide a short dictionary: our maps $\cR^0$ and $\cR$
reproduce the maps $\Phi$ and $\Phi^\sharp$ that feature in diagrams (1.67) and (1.74) in \cite{RIMS95},
respectively. The content of our Remark 5.3 is consistent with Theorem 5.7 in \cite{RIMS95}.},
simply because the embedding
$\Z_x:\hat P \to \hat P_c$ occurs in \cite{RIMS95} as well and its image is dense in $\hat P_c$.

To conclude,  we have shown in this paper that
the duality symplectomorphism between the system $(P,\omega,L)$ (\ref{1.1})
and the completion of the  system  $(\hat P, \hat \omega, \hat L)$ (\ref{1.3})
is nothing but the geometrically natural map between two models of
the reduced phase space
associated with  a symplectic reduction of the Heisenberg double of $U(n)$.
The character of the completion  of  the phase space  $(\hat P, \hat \omega)$
is thereby  illuminated:
$(\hat P, \hat \omega)$ represents a dense open submanifold of
the full reduced phase space wherein the reduced flows inherited from
the free flows on the double  are naturally complete.
We have also seen that the reduction turns
the two Abelian algebras $\H$ and $\hat\H$  (\ref{2.26})
(spanned by the   unreduced  free Hamiltonians)
 into the Abelian algebras of the
action and the particle-position variables of the system
$(P, \omega, L)$, respectively, and the r\^ole of these algebras is exchanged when viewed from the
perspective of the dual system.
In addition, we obtained the Ruijsenaars-Schneider Lax matrices
as well as  the integration algorithms for their flows
as easy by-products.
The present article, together with the previous one \cite{FKinJPA}, leave little doubt that
all
cases of the duality studied in \cite{SR-CMP,RIMS94,RIMS95} must permit analogous interpretation
in terms of symplectic reduction of  finite-dimensional
integrable systems of group theoretic
origin\footnote{The question of possible dualities in the elliptic case is wide open.
For the reduction approach to elliptic systems, see for example \cite{Arut} and
references therein.}.
The details are not trivial and we plan to return to   other
examples elsewhere.

It is proper to mention at this point that although
we expect that all cases of Ruijsenaars' duality
can be treated by the reduction method, the direct approach also
has its own advantages. For instance, it appears that
the detailed analyses of the scattering behaviour performed
in \cite{SR-CMP,RIMS94,RIMS95} cannot be
simplified by group theoretic means.

It can be surmised from results in \cite{FehPuszt,OblStok} that the classical trigonometric
$BC_n$ systems of van Diejen \cite{vD1} must admit a derivation based on  a reduction of the Heisenberg
double of $U(2n)$, which should produce the so far missing Lax matrices for these systems.
We plan to elaborate this, building on the description
of the  free systems given  in Section 2 of the present paper.
The results collected there might prove to be useful in other studies in
the reduction approach too, for example to obtain spin Ruijsenaars-Schneider systems.

Finally, it is pertinent  to remark that
 the  quantum mechanical analogue of Ruijsenaars' duality is the so-called bispectral
property \cite{DG} and it
is known that the quantum mechanical variants of the systems
$(P, \omega, L)$ and $(\hat P, \hat \omega,\hat L)$ form a
bispectral pair \cite{SR-Kup}. This follows from explicit inspection of the eigenfunctions
of the respective Hamiltonian operators. The eigenfunctions are provided by
Macdonald polynomials \cite{SR-Kup,vD2,Chalykh},
which also have close connections to quantum groups \cite{EK,Noum,Mim} and
to double affine Hecke algebras \cite{Cher,Obl}.
It would be
interesting to understand the bispectral property, at least for certain (integer) values of the coupling constant,
 in terms of a quantum analogue of
  our classical reduction.

                                                                            %
\renewcommand{\theequation}{\arabic{section}.\arabic{equation}}             %
\renewcommand{\thesection}{\Alph{section}}                                  %
\setcounter{section}{0}                                                     %
                                                                            %

\section{  Regularity of the moment map value $\nu(x)$}
\setcounter{equation}{0}
\renewcommand{\theequation}{A.\arabic{equation}}

The purpose of this appendix
is to demonstrate that $\nu(x)$ (\ref{bx}) is a regular value of the
Poisson-Lie moment map $\Lambda$ associated with the quasi-adjoint action on $GL(n,\bC)$.

We first recall from \cite{LMP} that every element $K\in GL(n,\bC)$ can be represented as
\be
K=g \triangleright  (bT^{-1}) \quad\hbox{with}\quad g\in G=U(n),\, T\in \bT_n, \, b\in B.
\label{A1}\ee
It was also shown in \cite{LMP} that every element of
the constraint-manifold
\be
F_{\nu(x)} = \{ K \in GL(n,\bC)\,\vert\, \Lambda(K)=\nu(x)\}
\label{A2}\ee
can be written as
\be
K = g \triangleright (\cN(T) a T^{-1})
\quad
\hbox{with}\quad g\in G_{\nu(x)},\, T\in \cC,\, a\in A.
\label{A3}\ee
Here $\cC$ is a subset of the regular elements $\bT_n^0 \subset \bT_n$, such that $\cC$  intersects
every  orbit of the permutation group $S(n)$  in $\bT_n^0$ precisely in one point.
$A<B$ is the group  of diagonal matrices with positive real entries, and
$\cN(T)$ belongs to the subgroup $N < B$ of upper-triangular
matrices with unit diagonal, given explicitly by the formula (\ref{N.1}).

The formula  (\ref{2.22}) of $\Lambda$ gives
\be
 \Lambda(bT^{-1})=bTb^{-1}T^{-1}\in N,
 \qquad
 \forall T\in \bT_n,\, \forall b\in B.
\label{A5}\ee
By using this together with
the $U(n)$ equivariance of $\Lambda$ and (\ref{A1}), we easily obtain that $\det \Lambda(K)=1$
for all $K\in GL(n,\bC)$.
In other words, $\Lambda$ takes its values in the normal subgroup $SB$ of $B$ consisting of
  elements of unit determinant.

\medskip
\noindent
{\bf Proposition A.1.}
\emph{The constant $\nu(x)$ (\ref{bx}) is a regular value of the Poisson-Lie moment map
$\Lambda: GL(n,\bC) \to SB$. }

\medskip
\noindent
\textbf{Proof.}
By the equivariance property of $\Lambda$ and (\ref{A3}),  it is sufficient to show
that the derivative (tangent) map
\be
\lambda_K:= T\Lambda(K): T_K GL(n,\bC) \to T_{\nu(x)} SB
\label{A6}\ee
is surjective
at every point of $F_{\nu(x)}$ of the form
\be
K= \cN(T) a T^{-1}
\quad
\hbox{with}\quad T\in \cC,\, a\in A.
\label{A7}\ee
Now the tangent space to $SB$ at any of its elements can be identified,
as a vector space, with $s\B$: the space of upper triangular, traceless
matrices with real diagonal entries.
This simply follows from the structure of $SB$ as a matrix group.
Take an arbitrary $Y\in s\B$ and consider the tangent vector
\be
Y K \in T_{K} GL(n,\bC),
\label{A8}\ee
which is the velocity of the curve $e^{Yt} K$ at $t=0$.
At $K$ of the form (\ref{A7}),
the formula (\ref{A5}) leads to
\be
\lambda_K( Y K) = Y \nu(x) - \nu(x) (T Y T^{-1})\in \mathrm{Lie}(N) \subset s\B,
\qquad
\forall Y\in s\B.
\label{A9}\ee
This equation implies that
for any $Z\in \mathrm{Lie}(N)$ there exists a unique $Y\in \mathrm{Lie}(N)$ for which
$\lambda_K(Y  K) = Z$.
To verify this, consider the principal gradation of $gl(n,\bC)$ by `heights' and
write accordingly
\be
Y= Y_1 + Y_2 +\cdots + Y_{n-1},
\quad
Z= Z_1 + Z_2 +\cdots + Z_{n-1},
\quad
\nu(x) = \1_n + \nu_1 + \nu_2 + \cdots + \nu_{n-1}.
\label{A10}\ee
First, the grade $1$ part of
\be
Y \nu(x) - \nu(x) (T Y T^{-1}) = Z
\label{A11}\ee
reads
\be
Y_1 - T Y_1 T^{-1} = Z_1,
\label{A12}\ee
which determines $Y_1$ uniquely in terms of $Z_1$ and $T$ since $T$ is regular.
Second, the grade $2$ part of (\ref{A11}) reads
\be
Y_2 - T Y_2 T^{-1} + ( Y_1 \nu_1 - \nu_1 T Y_1 T^{-1}) = Z_2,
\label{A13}\ee
which determines $Y_2$ uniquely in terms of $Z_ 1$, $Z_2$ and $T$.
Obviously, this argument can be continued increasing the grade until
the top grade, $(n-1)$, is reached.

We have seen  that the image of $\lambda_K$ (\ref{A6}),
at $K$ in (\ref{A7}),
  contains $\mathrm{Lie}(N)$.
Denoting by  $s\A$ the space of  diagonal elements of $s\B$,
we have
\be
s\B = s\A + \mathrm{Lie}(N),
\label{A14}\ee
and it remains to show that the $s\A$-projections of the elements
in the image of $\lambda_K$ span $s\A$.
For $W\in u(n)$,  denote by
$W \triangleright K$
the velocity of the curve $e^{tW} \triangleright K$ at $t=0$.
Because the moment map is equivariant, we obtain
\be
\lambda_K(W \triangleright K)= \mathrm{dress}_W \nu(x) \equiv
\nu(x) (\nu^{-1}(x) W \nu(x))_{\B},
\label{A15}\ee
where we use the unique decomposition
\be
X = X_{\B} + X_{u(n)},
\qquad
\forall X\in gl(n,\bC).
\label{A16}\ee
Note from (\ref{bx}) that
\be
\nu(x) =\1_n + ( 2 \sinh\frac{x}{2}) I_1 +\cdots,
\quad\hbox{where}\quad I_1= \sum_{j=1}^{n-1} E_{j,j+1}
\label{A18}\ee
and the dots stand for the higher grade parts of $\nu(x)$.
Let us now take an element $W\in u(n)$ of the form
\be
W= W_{-1} - (W_{-1})^\dagger,
\qquad
W_{-1}= \sum_{k=1}^{n-1} \xi_k E_{k+1,k}
\label{A19}\ee
with some \emph{real} numbers $\xi_k$.
Then we easily obtain
\be
\left(\nu^{-1}(x) W \nu(x))\right)_{\B} =
- \left(2 \sinh\frac{x}{2}\right) [I_1, W_{-1}] +
\hbox{non-zero grade contributions}.
\label{A20}\ee
Therefore, from (\ref{A14}), in this case we get the $s\A$-projection
\be
\pi_{s\A} \left(\lambda_K(W \triangleright K)\right) =
- 2 \left(\sinh\frac{x}{2}\right)\sum_{k=1}^{n-1} \xi_k (E_{k,k} - E_{k+1,k+1}).
\label{A21}\ee
These elements span $s\A$.  Hence we conclude that $\lambda_K$ (\ref{A6}) is surjective. \emph{Q.E.D.}

\medskip
\noindent
{\bf Remark A.1.}
 It was shown in \cite{LMP} (page 135) that the `constituents' $T\in \cC$, $a\in A$ and
$g\in G_{\nu(x)}$ that appear in (\ref{A3}) are uniquely determined by $K\in F_{\nu(x)}$ up to the
obvious freedom of multiplying $g$ by an element from the central $U(1)$ subgroup of $U(n)$.
This means, in particular, that $\bar G_{\nu(x)}$ (\ref{Gv}) acts freely on $F_{\nu(x)}$.
To avoid possible confusion, we point out that the set $\cC$ cannot be taken as a globally
smooth submanifold of $\bT_n^0$.  This follows from the non-triviality of the $S(n)$-bundle
$\bT_n^0\to Q(n)$; see the subsequent appendix.

\section{The structure of the configuration space $Q(n)$}
\setcounter{equation}{0}
\renewcommand{\theequation}{B.\arabic{equation}}

In this appendix we deal with  some features   of the   non-trivial manifold,
\be
Q(n)= \bT_n^0/S(n),
\label{S1}\ee
which is the configuration space of $n$ indistinguishable non-coinciding `point-particles'
moving on the circle.
We  shall expound the following statements:

\begin{enumerate}

\item{The principal $S(n)$-bundle $\bT_n^0 \to Q(n)$ is topologically non-trivial.}

\item{The manifold $Q(n)$ is orientable if and only if $n$ is odd.}

\item{$Q(n)$ cannot be separated into the Cartesian product of a `center of mass circle'
and a `space of relative positions'.}

\end{enumerate}

Note that the first statement  can be extracted also from  \cite{RIMS95} and
 the second one   generalizes the result of
Leinaas and
Myrheim  \cite{LM} who  showed that $Q(2)$ is the open M\"obius strip.

\medskip

To begin, we remark  that  the elements of the manifold $\bT_n^0$ can be viewed as the configurations of $n$
 distinguished non-coinciding points on a circle.
We can label those points by integers $1,...,n$ according to their registration on the diagonal in the matrix realization
of $\bT_n^0$. Consider  an element $E$ of  $\bT_n^0$ and its $S(n)$-action image $\sigma(E)$ under any permutation
$\sigma\in S(n)$ that changes
the cyclic order of the distinguished points on the circle.  It is evident
that we cannot smoothly connect the element $E$ with its image $\sigma(E)$  without violating the condition of non-coincidence.
This means that the element $E$  and its image  $\sigma(E)$ live respectively  in two different connected components of $T_n^0$.
On the other hand, it is not difficult to see  that  the permutations  respecting the cyclic order
(i.e.~the cyclic permutations)  can be realized smoothly, which means that the element $E$ and its image under the
cyclic permutation  live on the same connected component in $\bT_n^0$. Thus
 the connected components of $\bT_n^0$ correspond to
the coset space $S(n)/\bZ_n$. In particular, this implies that
the principal $S(n)$-bundle $\bT_n^0 \to Q(n)$ is topologically non-trivial, since if it were trivial then
the element $E$ and its image $\sigma(E)$ would live on different connected components for each permutation $\sigma$
not equal to the identity.

\medskip

Consider the  connected component $K$ of $\bT_n^0$ characterized by the requirement that the order of the
labels $1<2<...<n$
conforms with the cyclic order in which the distinguished points appear on the circle.
We know from the preceding paragraph that the manifold $K$ is  a topologically
non-trivial $\bZ_n$-covering of the manifold $Q(n)$.
In fact, the non-triviality of the $\bZ_n$-bundle $K \to Q(n)$ follows from the connectedness of $K$,
which also implies that $Q(n)$ is connected.
Now we remark that $K$ can be identified as the Cartesian product of the unit circle, on which the
distinguished points live, and of the $(n-1)$-dimensional open simplex $\operatorname{Simp}_{n-1}$ given by
\be
\operatorname{Simp}_{n-1}:= \{ \delta \in \bR^{n-1}\,\vert\,
\delta_j>0, \quad
\sum_{j=1}^{n-1} \delta_j < 2\pi\}.
\ee
Indeed, the point $z$ on the  unit circle corresponds to the position of the distinguished point   $1$,
the number $0<\delta_1<2\pi$ is the angle between the points $1$ and $2$, $\delta_2$ between the points $2$
and $3$ etc.  The generator $\Gamma$ of the cyclic
 group $\bZ_n$ is the cyclic permutation $(123...n)\to (n12...n-1)$. It is very easy to find the  action
 of $\Gamma$ on $K=U(1)\times \operatorname{Simp}_{n-1}$:
 \be
 \Gamma(z,\delta_1,\delta_2,...,\delta_{n-2},\delta_{n-1})=
 (ze^{\ri\delta_1},\delta_2,\delta_3,....,\delta_{n-1},2\pi-\sum_{j=1}^{n-1}\delta_j).
 \label{Pre}\ee
A simple  calculation of the Jacobian gives the transformation law of the volume form on
$K$ under the $\Gamma$-transformation:
\be
\Gamma^*(\ri z^{-1} dz \wedge  d\delta_1\wedge ...\wedge d\delta_{n-1})=
(-1)^{n+1}\ri z^{-1} dz\wedge  d\delta_1\wedge ...\wedge d\delta_{n-1}.
\ee
We observe that for $n$ odd  the volume form is $\bZ_n$ invariant and therefore it descends to an everywhere non-vanishing
$n$-form on the quotient $K/\bZ_n=Q(n)$. This means that for $n$ odd $Q(n)$ is orientable. Suppose now that there is also
an everywhere non-vanishing
$n$-form $\alpha$ on $Q(n)$ for $n$ even.  Its pull-back $\tilde\alpha$ to $K$ must be  given by a formula
\be
\tilde\alpha=f(z,\delta_1,...,\delta_{n-1}) \ri z^{-1} dz\wedge  d\delta_1\wedge ...\wedge d\delta_{n-1},\ee
where $f(z,\delta_1,...,\delta_{n-1})$ is a smooth and everywhere non-vanishing real function on $K$.
 On the other hand, it must hold also that
\be
\Gamma^*(\tilde\alpha)=\tilde\alpha,
\label{Sml}\ee
because otherwise $\tilde\alpha$ would not be the pull-back of $\alpha$ from $Q(n)$.  For $n$ even, the condition (\ref{Sml})
says that
\be
f(ze^{\ri \delta_1},\delta_2,\delta_3,....,\delta_{n-1},2\pi-\sum_{j=1}^{n-1}\delta_j)
=-f(z,\delta_1,\delta_2,...,\delta_{n-2},\delta_{n-1}).
\ee
Since  $f$ changes sign if we move from a point on $K$ to its $\Gamma$ image and $f$ is smooth,
it must vanish somewhere on $K$,  which  is in contradiction with the orientability of $Q(n)$ for $n$ even.

So far we have proved the first two statements displayed at the beginning.
For the last  statement, we notice  that the permutation
 action (\ref{Pre}) of $\bZ_n$ on $K$ commutes with the obvious action of $U(1)$ on $K$ given by
 \be
 u(z,\delta_1,\delta_2,...,\delta_{n-2},\delta_{n-1})= (uz,\delta_1,\delta_2,...,\delta_{n-2},\delta_{n-1}),
 \quad u\in U(1).
 \label{obvi}\ee
 Therefore this $U(1)$ action descends to  $Q(n)$. In specific systems, e.g.~in the trigonometric
 Ruijsenaars-Schneider and Sutherland systems,
the $U(1)$ action on $Q(n)$ just described  can be interpreted as global rotation symmetry.
Suppose that one tries to separate $Q(n)$ into the Cartesian product of a `center of mass
circle' and some manifold of `relative positions', say $R(n-1)$.
A reasonable definition of such separation requires
  the existence of  a product representation,
$Q(n)\simeq U(1) \times R(n-1)$,  such that $U(1)$ acts only on $U(1)$ and not on $R(n-1)$.
One might contemplate actions $\cA_u$ of the global rotations on
 $U(1) \times R(n-1)$ defined by
\be
\cA_u(w, r): = (u^k w, r),
\quad
\forall (w,r) \in U(1) \times R(n-1),
\label{sep}\ee
where $k$ is a fixed, non-zero integer. The choice $k=1$ might appear the most natural, while
$k=n$ corresponds to taking the hypothetical center of mass  as the product of the $n$ points on the unit circle.
However, we now show  that neither of these separations of $Q(n)$  exists.
Indeed, if  we consider a point $(z,\delta_1,...,\delta_{n-1})$ in $K$
such that $\delta_j=\frac{2\pi}{n}$ for every $j$, then we obtain
 \be
 e^{\frac{2\pi \ri}{n}}(z, \frac{2\pi}{n},\frac{2\pi}{n},...,\frac{2\pi}{n})= (ze^{\frac{2\pi \ri}{n}},
 \frac{2\pi}{n},\frac{2\pi}{n},...,\frac{2\pi}{n})=
 \Gamma(z, \frac{2\pi}{n}, \frac{2\pi}{n},...,\frac{2\pi}{n}).
 \ee
 Hence the action of  $e^{\frac{2\pi \ri}{n}}\in U(1)$ leaves invariant the point of $Q(n)$  covered by
  $(z, \frac{2\pi}{n}, \frac{2\pi}{n},...,\frac{2\pi}{n})\in K$. On the other hand, the
  isotropy group of the generic elements of $Q(n)$  is trivial under the natural $U(1)$ action.
In contrast,  under the  `separated action' (\ref{sep}) all points have the same isotropy group, $\bZ_k$.
This contradiction implies the validity of our third statement.

\bigskip
\bigskip
\noindent{\bf Acknowledgements.}
We wish to thank Ian Marshall for useful comments on the manuscript.
L.F. was partially supported
by the Hungarian
Scientific Research Fund (OTKA) under the grant K 77400.

\end{document}